\documentclass[11pt]{article}
\pdfoutput=1
\usepackage[utf8]{inputenc}
\usepackage{dsfont}
\usepackage{amsfonts}
\usepackage{amsmath}
\allowdisplaybreaks[4]        
\usepackage{amssymb}
\usepackage{euscript}     
\usepackage{braket}
\usepackage{starfont}
\usepackage{color,soul}         
\usepackage{tensor}        
\usepackage{amsthm}
\usepackage{graphicx}
\usepackage{slashed}
\usepackage{leftidx}
\usepackage{wasysym}

\newcommand{\rmi}{\mathrm i}

\usepackage{mathtools}
\usepackage{graphicx}
\usepackage{float}
\usepackage{dsfont} 
\usepackage{cancel}
\usepackage{enumitem}
\usepackage[normalem]{ulem}

\usepackage{simplewick}
\usepackage{bbm}
\usepackage{bbold} 

\usepackage{hyperref}
\usepackage{physics, xparse}
\usepackage{subfigure}
\usepackage{bbm}

\definecolor{outerspace}{rgb}{0.25, 0.29, 0.3}
\definecolor{scarlet}{rgb}{1.0, 0.13, 0.0}
\usepackage[header,title,page,titletoc]{appendix}  
\definecolor{princetonorange}{rgb}{1.0, 0.56, 0.0}
\definecolor{WildStrawberry}{rgb}{1.0, 0.26, 0.64}
\definecolor{rossocorsa}{rgb}{0.83, 0.0, 0.0}
\definecolor{navyblue}{rgb}{0.0, 0.0, 0.5}
\usepackage[numbers,sort&compress]{natbib}  
\usepackage{float}
\usepackage[paper=letterpaper,margin=1in]{geometry}
\newcommand{\req}[1]{(\ref{#1})} 

\newcommand{\bea}{\begin{eqnarray}}
\newcommand{\diff}{\mathrm{d}}
\newcommand{\eea}{\end{eqnarray}}
\newcommand{\ba}{\begin{eqnarray}}
\newcommand{\ea}{\end{eqnarray}}

\newcommand{\be}{\begin{equation}}
\newcommand{\ee}{\end{equation} }
\newcommand{\beqa}{\begin{eqnarray}}
\newcommand{\eeqa}{\end{eqnarray}}
\newcommand{\beqar}{\begin{eqnarray*}}
\newcommand{\eeqar}{\end{eqnarray*}}

\renewcommand{\req}[1]{eq.~(\ref{#1})}
\newcommand{\rme}{\mathrm e}
\newcommand{\rmd}{\mathrm d}

\newcommand{\eg}{{\it e.g.,}\ }
\newcommand{\ie}{{\it i.e.,}\ }

\usepackage{hyperref}
\hypersetup{
    colorlinks,
    citecolor=rossocorsa,
    filecolor=navyblue,
    linkcolor=navyblue,
    urlcolor=navyblue
}

\begin{document} 

\begin{titlepage}

\begin{center}

\phantom{ }
\vspace{3cm}

{\bf \huge{Aspects of higher-curvature gravities\\ with covariant derivatives}}
\vskip 0.5cm
{\bf Sergio E. Aguilar-Gutierrez\,${}^{\text{\neptune}}$, Pablo Bueno\,${}^{\text{\jupiter}}$,
 Pablo A. Cano\,${}^{\text{\jupiter}}$,\\ Robie A. Hennigar\,${}^{\text{\jupiter}}$ and Quim Llorens\,${}^{\text{\jupiter}}$}
\vskip 0.04in
\small{  ${}^{\text{\neptune}}$\textit{Instituut voor Theoretische Fysica, KU Leuven,}}
\vskip -.4cm
\small{\textit{ Celestijnenlaan 200D, B-3001 Leuven, Belgium}}\normalsize
\vskip 0.05in
\small{
\vskip -0.04in
${}^{\text{\jupiter}}$\textit{Departament de F{\'\i}sica Qu\`antica i Astrof\'{\i}sica,  Institut de Ci\`encies del Cosmos\\
 Universitat de
Barcelona,~Mart\'{\i} i Franqu\`es 1, E-08028 Barcelona, Spain}}

\begin{abstract}

We study various aspects of higher-curvature theories of gravity built from contractions of the metric, the Riemann tensor and the covariant derivative, $\mathcal{L}(g^{ab},R_{abcd},\nabla_a)$. We characterise the linearized spectrum of these theories and compute the modified Newton potential in the general case. 
Then, we present the first examples of Generalized Quasi-topological (GQT) gravities involving covariant derivatives of the Riemann tensor. We argue that they always have second-order equations on maximally symmetric backgrounds. Focusing on four spacetime dimensions,  we find new densities of that type involving eight and ten derivatives of the metric. In the latter case, we find new modifications of the Schwarzschild black hole. These display thermodynamic properties which depart from the ones of polynomial GQT black holes. In particular, the relation between the temperature and the mass of small black holes, $T\sim M^{1/3}$, which universally holds for general polynomial GQT modifications of Einstein gravity, gets modified in the presence of the new density with covariant derivatives to $T\sim M^{3}$. Finally, we consider brane-world gravities induced by Einstein gravity in the AdS bulk. We show that the effective quadratic action for the brane-world theory involving arbitrary high-order terms in the action can be written explicitly in a closed form in terms of  Bessel functions. We use this result to compute the propagator of metric perturbations on the brane and its pole structure in various dimensions, always finding infinite towers of ghost modes, as well as tachyons and more exotic modes in some cases.

\end{abstract}
\end{center}
\end{titlepage}

\setcounter{tocdepth}{2}

{\parskip = .2\baselineskip \tableofcontents}


\section{Introduction}
\label{sec:Introduction}

Despite the spectacular list of experimental successes of general relativity, there are good reasons to explore alternatives to Einstein's theory. Firstly, it is expected that the Einstein-Hilbert action is the first in an infinite series of terms involving an increasing number of derivatives of the metric \cite{Endlich:2017tqa}. 
This can be seen explicitly within the string theory framework, where the new terms appear weighted by powers of the inverse string tension  \cite{Gross:1986mw,Green:1997tv,Frolov:2001xr}. Additionally, holographic higher-curvature gravities can be used, through AdS/CFT \cite{Maldacena,Witten}, as toy models of conformal field theories (CFTs) which, being inequivalent from their Einsteinian counterparts, can sometimes be used to unveil new universal properties valid for completely general CFTs \cite{Kats:2007mq,Brigante:2007nu,Buchel:2008ae,Myers:2010xs,Camanho:2010ru,Mezei:2014zla,Bueno1,Bueno:2018yzo,Bueno:2022jbl}.

From a different perspective, it is important to characterise the possible existence (or lack thereof) of universal features of classical gravity in regimes in which the Einsteinian description is expected to receive higher-curvature corrections \cite{PabloPablo4,Arciniega:2018tnn}. In order to do this, it is often convenient to consider particular classes of higher-curvature gravities displaying certain special properties. The list includes:  quadratic \cite{Alvarez-Gaume:2015rwa,Lu:2015cqa}, Lovelock \cite{Lovelock1,Lovelock2,Boulware:1985wk,Padmanabhan:2013xyr}, Quasi-topological \cite{Quasi2,Quasi,Dehghani:2011vu,Ahmed:2017jod,Cisterna:2017umf} and Generalized Quasi-topological gravities (GQTs) \cite{Hennigar:2017ego,PabloPablo3,Bueno:2022res,Moreno:2023rfl}, among others \cite{Lu,Karasu:2016ifk,Li:2017ncu,Bueno:2016dol}. All of these belong to the subset of theories built from contractions of the Riemann tensor and the metric. In particular, GQTs ---which are characterised by admitting ``single function'' static and spherically symmetric solutions (see Section~\ref{gqT}) as well as possessing second-order equations  on maximally symmetric backgrounds ---  have been shown to provide a basis for general gravitational effective actions built from general contractions of the Riemann tensor and the metric: any $\mathcal{L}(g^{ab},R_{abcd})$ theory can be mapped order by order, via a field redefinition, to certain GQT \cite{Bueno:2019ltp}.

 Although seemingly less likely, it is also possible that deviations from Einstein gravity are eventually measured in unexpected situations (\eg beyond the effective field theory regime) and it is important to have alternative predictions which can be tested \cite{Cayuso:2023xbc}.  Along this direction there have been numerous attempts at constructing alternatives to general relativity which are compatible with all current observations and internally consistent. This includes again quadratic theories \cite{Stelle:1976gc,Salvio:2018crh,Salvio:2019ewf}, $f(R)$ models \cite{Sotiriou}, as well as non-local gravities which, by including an infinite number of derivatives in the action, can be made free of ghosts \cite{Tomboulis:1997gg,Biswas:2005qr,Biswas:2010zk,Biswas:2011ar,Modesto:2011kw,Biswas:2013cha,Modesto:2014lga,Frolov:2015bta}.  
 Non-local gravities are particular instances of the general set of theories which will be subject of study in the present paper, namely, diffeomorphism-invariant theories constructed from general contractions of the Riemann tensor and its covariant derivatives,
\begin{equation}\label{GeneralAction}
I=\frac{1}{16\pi G} \int \diff^D x \sqrt{|g|}\, \mathcal{L}( g^{ab},R_{abcd},\nabla_a)\, .
\end{equation}
As a matter of fact, terms involving covariant derivatives of the Riemann tensor generically appear in gravitational effective actions \cite{Ruhdorfer:2019qmk,Li:2023wdz}. A scenario in which this is apparent corresponds to the so-called brane-world gravities \cite{Randall:1999ee,Randall:1999vf,Karch:2000ct}. These are effective gravitational theories defined on the world volume of branes inserted on higher-dimensional spacetimes.  Originally introduced with phenomenological motivations, they have received a lot of attention recently in the holographic context ---see \eg  \cite{Chen:2020uac,Chen:2020hmv,Emparan:2020znc}.


In this paper we present the first examples of GQT gravities with covariant derivatives. Analogously to their ``polynomial'' counterparts, we show that they have second-order linearized equations on maximally symmetric backgrounds\footnote{This provides a counterexample to the conjecture of \cite{Edelstein:2022lco} regarding the absence of theories with covariant derivatives of the curvature possessing an Einsteinian spectrum.} and that they admit black hole solutions characterized by a single function, $g_{tt} g_{rr}=-1$. Focusing on four dimensions, we find that the lowest-order instances of  GQT densities involve eight derivatives of the metric. However, we observe that all such theories admit the Schwarzschild metric as a solution, and therefore do not give rise to new solutions when considered as corrections to general relativity. The first GQT density with covariant derivatives which does correct the Schwarzschild solution occurs at tenth order in derivatives of the metric ---see \req{L1049} below for its explicit form. For this, we find that the corrected solution displays features similar to the ones of polynomial GQTs, including a near-horizon expansion fully determined by a single parameter to be fixed asymptotically or the possibility of determining their thermodynamic properties in a fully analytic fashion for general values of the coupling. We analyse some of the thermodynamic properties of the new solution finding, in particular, that for small black holes the temperature behaves as a function of the mass as $T \sim M^{3}$ as $r_h \to 0$. This departs from the behaviour encountered for all polynomial GQT theories, for which it has been argued that the relation $T \sim M^{1/3}$ holds universally for small black holes \cite{PabloPablo4, Bueno:2019ycr}. 

The analysis of the linearized spectrum of GQTs is performed after obtaining some general results on the linearization of general higher-curvature theories with covariant derivatives. We present general formulas which allow for the computation of the linearized equations of a given higher-curvature theory from its effective quadratic action. Using this, we show that both GQTs as well as brane-world gravities belong to the family of theories which do not include scalar modes in their linearized spectrum. Additionally, we obtain a formula for the modified Newton potential valid for general higher-curvature theories involving covariant derivatives in arbitrary dimensions.


Then, we move to brane-world theories. The effective gravitational action induced on the brane world volume is given by an infinite series of higher-derivative terms of the form \cite{Kraus:1999di,Emparan:1999pm,Balasubramanian:1999re,Papadimitriou:2004ap,Papadimitriou:2010as,Elvang:2016tzz,Bueno:2022log,Anastasiou:2020zwc}
\begin{equation}\notag
I_{\rm bgrav}=\int \frac{ \diff^{D}x\sqrt{-g}}{16\pi G_{D}} \left[  R +\frac{\ell^2}{(D-2)(D-4)} R^{ab}\left( R_{ab}-\frac{D}{4(D-1)}g_{ab}R \right) +\dots\right]\, ,\label{totalactiondd}
\end{equation}
where $\ell$ is the AdS$_{D+1}$ radius of the ambient spacetime. Starting at sixth order in derivatives, all the higher-curvature densities involve terms with covariant derivatives of the Riemann tensor ---\eg see \req{cubi} for the next order. Here we show that the effective curvature-squared action of the full brane-world gravity ---including the infinite tower of terms with covariant derivatives --- can be written as
\begin{align}\notag
I_{\rm bgrav}^{(2)}=\int \frac{\diff^{D}x\sqrt{-g}}{16\pi G_{D}}  \left[R+\ell^2 R^{ab}F_D \left(\ell^2\Box\right) \left( R_{ab}-\frac{D}{4(D-1)} g_{ab }R\right)\right]\, ,\label{totalaction}
\end{align}
where
\begin{equation}
F_{D}(\ell^2\Box)\equiv \frac{D(D-2)}{\ell^4\Box^2}-\frac{1}{\ell^2\Box}-\frac{(D-2)Y_{\frac{D+2}{2}}\left(\ell \sqrt{\Box}\right)}{\ell^3 \Box^{3/2}Y_{\frac{D}{2}}\left(\ell \sqrt{\Box}\right)}\, ,
\end{equation}
and $Y_{k}$ are Bessel functions of the second kind. Using this expression, we study the linearized spectrum of  the theory on Minkowski spacetime in various dimensions. Generically,  the metric perturbations propagator includes poles of the form
\begin{equation}
P_D(\ell^2 k^2) \sim \frac{1}{\ell^2 k^2}\, ,\qquad P_D(\ell^2 k^2) \sim -\frac{2}{(D-2)\ell^2 [k^2+m_j^2]} \, ,
\end{equation}
where the first is the usual Einstein gravity massless spin-2 mode, and the second corresponds to infinite towers of massive spin-2 modes (labelled by $j$) which always have negative kinetic energy. Depending on the dimension, some of those modes have positive squared masses, some of them have negative squared masses and some of them have imaginary squared masses.

The structure of the paper is the following. Section \ref{lini} contains some comments on the structure of the linearized equations of  general higher-curvature gravities with covariant derivatives on general maximally symmetric backgrounds, a characterization of the structure of poles of the metric propagator on Minkowksi spacetime as well as a derivation of the generalized Newton potential. In Section \ref{gqT} we construct GQTs with covariant derivatives in four spacetime dimensions and study their new black-hole solutions and their thermodynamic properties. In Section \ref{bW} we study the linearization of brane-world gravities obtaining their effective quadratic action and characterizing the pole structure of the metric propagator in various dimensions. We conclude in Section \ref{conclu} with some comments on future directions. Appendix \ref{basis} contains a complete list of the curvature invariants at each order in derivatives up to eight, as well as the non-exhaustive set we have used at order ten. In Appendix \ref{8dpure} we  present new hairy black hole solutions of pure eight-derivative GQTs with covariant derivatives. Finally, in Appendix \ref{adsbck} we present the linearized field equations around an AdS background for the simplest examples of the theories we consider here.

\section{Linearized higher-curvature gravities with covariant derivatives}\label{lini}
Throughout the paper we will be interested in the linearized equations of various higher-curvature theories with covariant derivatives. In this section we analyse the structure of such equations for a general theory of the form (\ref{GeneralAction}) in general dimensions.
 We derive their general form on a maximally symmetric background and then, focusing on the Minkowski case, we identify the precise relation between the effective quadratic action and the linearized equations, classifying the different theories according to the modes propagated. In particular, we identify a set of generalisations of a particular type of quadratic densities involved in the definition of the so-called ``critical gravities'' ---which have the peculiarity of propagating no scalar modes. This set of theories will include both the new GQTs theories presented in Section \ref{gqT} and the brane-world theories studied in Section \ref{bW} as particular instances. Additionally, we obtain an explicit formula for the $D$-dimensional generalized Newton potential resulting from a general higher-curvature gravity with covariant derivatives. 
 
 Before starting, let us point out that many of the results presented in this section have appeared in different forms in previous literature. Indeed, both the linearization on maximally symmetric backgrounds of general $\mathcal{L}(g^{ab},R_{abcd},\nabla_a)$ theories as well as the Newton potential have been studied in the four-dimensional case in \cite{Biswas:2013kla,Conroy:2014eja,Biswas:2016etb,Edholm:2018wjh,Biswas:2016egy,Dengiz:2020xbu, Kolar:2023mkw}. 

We are interested in gravity theories of the form (\ref{GeneralAction}).
Sometimes it is convenient to split the Lagrangian as follows
\begin{equation}
 \mathcal{L}( g^{ab},R_{abcd},\nabla_a)=\frac{(D-1)(D-2)}{\ell^2}+R+ \mathcal{L}_{\rm R}( g^{ab},R_{abcd})+ \mathcal{L}_{\nabla}( g^{ab},R_{abcd},\nabla_a)\, ,
\end{equation}
where we included an explicit Einstein-Hilbert plus (negative) cosmological constant piece, $\mathcal{L}_{\rm R}$ includes terms which do not involve covariant derivatives, and $\mathcal{L}_{\nabla}$ includes terms which contain at least one covariant derivative of the Riemann tensor. 
%
The equations of motion for this theory can be written as \cite{Iyer:1994ys}
\begin{equation}
\mathcal{E}_{ab}\equiv T_{a}\,^{cde}R_{b cde}-\frac{1}{2}g_{ab}\mathcal{L}-2 \nabla^c\nabla^d T_{a c d b}=0
\end{equation}
where
\begin{equation}
T^{abcd}\equiv \left[ \frac{\partial \mathcal{L}}{\partial R_{abcd}} - \nabla_{a_1} \frac{\partial \mathcal{L}_{\nabla}}{\partial \nabla_{a_1} R_{abcd}}+\dots+ (-1)^m \nabla_{(a_1}\dots \nabla_{a_m)}\frac{\partial \mathcal{L}_{\nabla}}{\partial  \nabla_{(a_1}\dots \nabla_{a_m)}R_{abcd}} \right]\, .
\end{equation}
In the case of maximally symmetric backgrounds with metric $\bar g_{ab}$, the Riemann tensor is given by
\begin{equation}
\bar R_{abcd}= -\frac{2}{\ell_\star^2} \bar g_{a[c}\bar g_{d]b}\, ,
\end{equation}
where $\ell_\star^2$ has dimensions of length$^2$ and it is a positive number in the case of an AdS$_D$ background, a negative number in the case of dS$_D$, and infinite for Minkowski.  In order for $\bar g_{ab}$ to be a solution of $\mathcal{L}( g^{ab},R_{abcd},\nabla_a)$, the equations of motion impose the algebraic equation \cite{Aspects} 
\begin{equation}
1-\chi+\frac{\ell^2}{(D-1)(D-2)}\left[\mathcal{L}_{\rm R}(\chi)-\frac{2\chi}{D}\mathcal{L}_{\rm R}'(\chi)\right]=0\, ,
\end{equation}
 where we defined $\chi \equiv \ell^2/\ell_{\star}^2$,  $\mathcal{L}_{\rm R}(\chi)$ stands for the on-shell evaluation of the corresponding Lagrangian on the maximally symmetric background, and $\mathcal{L}_{\rm R}'(\chi)\equiv \diff \mathcal{L}_{\rm R}(\chi)/\diff \chi$. Observe that the piece of the Lagrangian involving covariant derivatives of the Riemann tensor makes no contribution to this equation, which follows from $\bar \nabla_a \bar g_{bc}=0$. Naturally, for Einstein gravity the above equation simply imposes the condition $\chi=1$. For a Lagrangian built from polynomials of the Riemann tensor  involving densities up to order $n$ in the curvature, the above equation is an order-$n$ algebraic equation for $\chi$, which will in general have many possible solutions, depending on the values of the corresponding higher-derivative couplings.

\subsection{Linearized equations}

Let us now consider the linearized equations of a general theory of the form given by \req{GeneralAction} around a maximally symmetric background. We expand the metric as
\begin{equation}
g_{ab}=\bar g_{ab}+  h_{ab}\, ,
\end{equation}
where $h_{ab}$ is a small perturbation. Every relevant object built from the metric can then be expanded at the desired order in the pertubation as $T=T^{(0)}+T^{(1)}+T^{(2)}+\mathcal{O}(h^3)$. 

Given a particular theory we have two routes to derive its linearized equations. On the one hand, we can  take the full non-linear equations and expand each of the terms to linear order in the perturbation. Alternatively, we can expand the action to second order in the perturbation and derive the linearized equations from the first variation. 
As we have seen, the full non-linear equations of a theory like  (\ref{GeneralAction}) have a rather complicated form. However, it is not difficult to argue that the most general form of the linearized equations is much simpler. In order to see this, let us start by characterizing all possible terms that may arise in the linearized equations. Doing this amounts to classify all symmetric tensors of 2 indices built from $R_{abcd}^{(1)}$, $\bar g_{ab}$ and $\bar \nabla_a$ which are linear in the metric perturbation. 

Let us start with a few observations. First, observe that the linearized Riemann tensor is linear in $h_{ab}$, and therefore all possible terms will have a single Riemann tensor, possibly acted upon with covariant derivatives and with various indices contracted. 
Another observation is that all terms must necessarily contain an even number of covariant derivatives, since $\nabla_a$ is the only available object with an odd number of indices.  In addition, note that all Riemann tensors will actually appear in the form of Ricci tensors. This is because: a) any term involving exclusively metrics and Riemann tensors reduces to Ricci tensors or vanishes, since at most two of the indices can remain uncontracted; b) any term involving covariant derivatives and Riemann tensors reduces to covariant derivatives and Ricci tensors. Indeed,  when only two indices are left uncontracted, a tensor of the form
\begin{equation}
\nabla_a\nabla_b R_{cdef}\, ,
\end{equation}
reduces to one of the following four possibilities: $\nabla^c\nabla^e R_{cdef},\, \Box R_{ab},\, \nabla_a \nabla_b R,\, 0 $. In addition, using the second Bianchi identity it follows that the first possibility can only give rise to a linear combination of the second and the third, plus higher-order terms in $h$. 
We therefore conclude that the most general possible term will come from contracting all but two indices in an expression of the form
\begin{equation}
\Box^l \nabla_{c_1}\nabla^{c_2} \dots \nabla_{c_{2m-1}}\nabla^{c_{2m}} R_{ab}\,,
\end{equation}
where $c_i \neq c_j \, \forall i\neq j$. Contracting $2m$ of the indices, we immediately see that the only three possibilities are in fact
\begin{equation}
\bar g_{ab} \Box^l R\, , \quad \nabla_a\nabla_b \Box^l R\, , \quad \Box^lR_{ab}\, .
\end{equation}
We then conclude that the linearized equations of a general $\mathcal{L}(g^{ab},R_{abcd},\nabla_a)$ theory around maximally symmetric backgrounds will always take the form
\begin{equation}\label{eq:Eab general msb}
\mathcal{E}_{ab}\equiv \sum_{l=0}\ell^{2l} \left[ \alpha_l  \bar \Box^l G^{(1)}_{ab}+ \beta_l  \bar \Box^l R^{(1)} \bar g_{ab} +\gamma_{l+1} \ell^2 \Box^l [\bar g_{ab}\bar \Box- \bar\nabla_a\bar\nabla_b]  R^{(1)} \right] = 0\, ,
\end{equation}
for certain dimensionless constants $\alpha_l,\beta_l,\gamma_l$ which will be related to the gravitational couplings, and where we rearranged some of the terms for later convenience. Implicitly, we have assumed that the theory involves a polynomial dependence on the covariant derivatives. Relaxing this requirement, would yield the more general form
\begin{equation}\label{eq:Eab general msb2}
\mathcal{E}_{ab}\equiv \left[ f_1(\ell^2 \bar \Box) G^{(1)}_{ab}+  f_2(\ell^2 \bar \Box) R^{(1)} \bar g_{ab} +   f_3(\ell^2 \bar \Box)  [\bar g_{ab}\bar \Box- \bar\nabla_a\bar\nabla_b]R^{(1)} \right] = 0\, ,
\end{equation}
for certain functions $f_1, f_2, f_3$. The form of the equations can be further constrained by noting that the tensor $\mathcal{E}_{ab}$ must be divergence-free, that is, $\bar\nabla^{a}\mathcal{E}_{ab}=0$. By commuting $\bar{\nabla^{a}}$ and $\bar\Box$, one can show that the divergence reads
\begin{equation}
\begin{aligned}
\bar\nabla^{a}\mathcal{E}_{ab}=\Bigg\{&f_2\left(\ell^2\tilde \Box\right)+\frac{D-1}{D}\left[f_{3}\left(\ell^2\tilde \Box\right)\tilde\Box-f_{3}\left(\ell^2\hat \Box\right)\left(\bar\Box-\frac{1}{\ell_{\star}^2}\right)\right]\\
&+\frac{2-D}{2D}\left[f_{1}\left(\ell^2\tilde \Box\right)-f_{1}\left(\ell^2\hat\Box\right)\right]\Bigg\}\bar\nabla_{b}R^{(1)}\, ,
\end{aligned}
\end{equation}
where
\begin{equation}
\hat\Box=\bar\Box-\frac{D+1}{\ell_{\star}^2}\, ,\quad \tilde\Box=\bar\Box+\frac{D-1}{\ell_{\star}^2}\, .
\end{equation}
Therefore, the function $f_2$ is not free, but it depends on $f_1$ and $f_3$ by
\begin{equation}\label{f2relation}
\begin{aligned}
f_2\left(\ell^2\bar\Box\right)&=\frac{D-2}{2D}\Big[f_1\left(\ell^2\bar\Box\right)-f_1\left(\ell^2\bar\Box+2D\chi\right)\Big]
\\
&+\frac{D-1}{D\ell^2}\Big[f_3\left(\ell^2\bar\Box+2D \chi\right)\left(\ell^2\bar\Box-\chi D\right)-f_3\left(\ell^2\bar\Box\right)
\ell^2\bar\Box\Big]\, ,
\end{aligned}
\end{equation}
where we recall that $\chi=\ell^2/\ell_{\star}^2$. Observe that in the case of flat space, $f_2$ vanishes.


In the case of theories which do not involve covariant derivatives, it is known that the most general form of the linearized equations is captured by a general quadratic action in the Riemann tensor. Something similar happens for a general $\mathcal{L}(g^{ab},R_{abcd},\nabla_a)$ theory. Indeed, in that case the most general quadratic action reads
\be \label{effq}
\mathcal{L}_{\rm eff} = \lambda\left[\frac{(D-1)(D-2)}{\ell^2}+R+ \ell^2 R F_1(\ell^2 \bar \Box) R + \ell^2 R_{ab}   F_2(\ell^2 \bar \Box) R^{ab}+\ell^2 R_{abcd} F_3(\ell^2\bar \Box)R^{abcd}\right]\, ,
\ee
for certain functions $F_1,F_2,F_3$. It is then possible to relate these to the functions $f_{1}$ , $f_{2}$ , $f_{3}$ of the linearized equations \eqref{eq:Eab general msb2}. Such relation turns out to be quite cumbersome in the case of (A)dS backgrounds, as we illustrate in the Appendix~\ref{adsbck}. In what follows we analyze the case of Minkowski backgrounds. 

\subsubsection{Minkowski background}\label{minko}
When the background is flat, the linearized equations for the quadratic Lagrangian \eqref{effq} read\footnote{Note that, as far as the linearized equations on Minkowski space are concerned,  the term $R^{abcd} \Box^n R_{abcd}$ is not independent from the other two. Indeed, one finds 
\begin{equation}
    R \Box^n R - 4 R^{ab} \Box^n R_{ab} + R^{abcd} \Box^n R_{abcd} = \text{Total Derivative} + \mathcal{O}(R_{abcd}^3) \,.
\end{equation}
Hence, in the Minkowski case we could have just redefined out $F_3$ in $\mathcal{L}_{\text{eff}}$ without loss of generality. }
\bea \notag
&&\frac{\lambda}{2} \left\{ \left[1+\left[4F_3(\ell^2 \bar \Box)+F_2(\ell^2 \bar \Box)\right] \ell^2\bar\Box \right] G_{ab}^{(1)} \right. \\ && \left.-\ell^2\left[ 2F_1(\ell^2 \bar \Box)+F_2(\ell^2 \bar \Box)+2F_3(\ell^2 \bar \Box)\right][ \bar\nabla_a\bar\nabla_b-\bar g_{ab}\bar \Box]R^{(1)} \right\} =0\, . \label{lineareqs}
\eea
where
\begin{align}
G_{ab}^{(1)}&=-\frac{1}{2}\bar\Box h_{ab}+\bar\nabla_{(a|}\bar\nabla^{c}h_{c|b)}-\frac{1}{2}\bar\nabla_{a}\bar\nabla_{b}h-\frac{1}{2}\bar g_{ab} R^{(1)}\, ,\\
R^{(1)}&=\bar\nabla^{a}\bar\nabla^{b}h_{ab}-\bar \Box h\, ,
\end{align}
are the linearized Einstein tensor and Ricci scalar, respectively. We point out that the above linearized equations can be obtained immediately using the result found in \cite{Aspects} for theories which do not involve covariant derivatives of the Riemann tensor. The idea is to use the same relations between the quadratic action couplings and the constant parameters ($a,b,c,e$) appearing in such equations but now promoting the constants to functions of $\ell^2\bar \Box$. 

 The trace of the equations reads
\be
-\frac{\lambda}{4} \left[(D-2)-\ell^2 \bar \Box  \left[4F_3(\ell^2 \bar \Box)+D F_2(\ell^2\bar \Box)+4(D-1)F_1(\ell^2 \bar\Box) \right] \right] R^{(1)}=0\, ,
\ee
and their traceless part is given by
\bea\notag
&&
\frac{\lambda}{2} \left\{  \left[1+\left[4F_3(\ell^2 \bar \Box)+F_2(\ell^2 \bar \Box)\right]\ell^2\bar\Box \right] R^{(1)}_{\langle a b\rangle }\right. \\ && \left.- \ell^2\left[ 2F_1(\ell^2 \bar \Box)+F_2(\ell^2 \bar \Box)+2F_3(\ell^2 \bar \Box)\right] \bar\nabla_{\langle a}\bar\nabla_{b\rangle}R^{(1)} \right\}=0\, .
\eea
Observe now that for theories satisfying the condition
%
\begin{equation}\label{keyko}
4F_3(\ell^2 \bar \Box)+D F_2(\ell^2\bar \Box)+4(D-1)F_1(\ell^2 \bar\Box)=0\, ,
\end{equation}
the trace equation becomes second order and simply reads
\begin{equation}\label{oop}
-\frac{\lambda}{4}(D-2) R^{(1)}=0\, ,
\end{equation}
which is nothing but the Einstein gravity result. In the case in which $F_i=\alpha_i$ are constants, condition (\ref{keyko}) selects a linear combination of quadratic terms which appear in the so-called ``critical gravities'' in general dimensions---see \eg \cite{Lu:2011zk,Maldacena:2011mk,Bergshoeff:2009hq,Oliva:2011xu,Hassan:2013pca,Kan:2013moa,Anastasiou:2016jix}. In particular, the action reduces in that case to
\begin{equation}\label{confi}
\mathcal{L}_{\rm eff}=\lambda \left[ \frac{(D-1)(D-2)}{\ell^2}+R + \ell^2 \alpha_3 \mathcal{X}_4 + \ell^2 (\alpha_1-\alpha_3) \left(R^2-\frac{4(D-1)}{D}R_{ab}R^{ab} \right)  \right]\, ,
\end{equation}
where $\mathcal{X}_4\equiv R^2-4R_{ab}R^{ab}+R_{abcd}R^{abcd}$ is the Gauss-Bonnet density and the second term can be written as a linear combination of  $\mathcal{X}_4$ and the Weyl tensor squared. For this theory, the linearized spectrum on a general maximally symmetric background is known to involve the usual massless graviton and the massive one, but not the scalar mode. This is also the case for theories satisfying \req{keyko} with non-constant functions. As we will see later, both Generalized Quasi-topological and Brane-world gravities belong to that class.

In order to study the physical modes propagated by the metric perturbation, let us now fix the harmonic gauge, which amounts to setting
\begin{equation}
\bar\nabla^a h_{ab}= \frac{1}{2}\bar\nabla_b h\, .
\end{equation} 
Then, the linearized Einstein tensor and Ricci scalar become
\begin{equation}
G_{ab}^{(1)}=-\frac{1}{2}\bar\Box h_{ab}+\frac{1}{4}\bar g_{ab} \bar\Box h\, , \qquad R^{(1)}=-\frac{1}{2}\bar\Box h\, .
\end{equation}
For theories satisfying \req{keyko}, the trace equation (\ref{oop}) imposes $\bar\Box h=0$. Using the residual gauge freedom $h_{ab}\rightarrow h_{ab}+\nabla_{(a}\xi_{b)}$, with $\Box\xi_a=0$, we can set $h=0$. Therefore, the trace of the perturbation has no dynamics and there are no scalar modes. On the other hand, the traceless part of the equations becomes
\bea\label{tracelesss}
-\frac{\lambda}{4} \left[1+\left[4F_3(\ell^2 \bar \Box)+F_2(\ell^2 \bar \Box)\right]\ell^2\bar\Box \right] \bar \Box h_{\langle a b \rangle}=0\, .
\eea
By performing the Fourier transform in this expression, which amounts to $\Box\rightarrow -k^2$, we can read off the propagator
\begin{equation}
P(k)=\frac{4}{\lambda k^2\left[1-   \ell^2 k^2 \left(4F_3(-\ell^2 k^2)+F_2(-\ell^2 k^2)\right)\right]}\, .
\end{equation}
Poles of the propagator inform about the degrees of freedom of the theory. For each pole, $k^2=-m^2$ indicates the mass. Thus, imaginary poles correspond to massive modes, while real poles are tachyonic modes. On the other hand, the residue of each pole tells us about the energy carried out by the corresponding mode. A positive residue ---like the massless graviton one, $k^2=0$--- corresponds to positive energy, and viceversa for a negative residue. For constant functions, $F_i=\alpha_i$, we have the poles
\begin{equation}
m^2=0\, , \quad m_g^2=-\frac{1}{(4\alpha_3+\alpha_2)\ell^2}\, ,
\end{equation}
corresponding to the anticipated massless and massive graviton, respectively, and in agreement with the result of \cite{Aspects,Sisman:2011gz}. The next to simplest case corresponds to
$F_i(\ell^2 \bar \Box)=\alpha_i+\beta_i \ell^2\bar \Box $. For that, one finds 
\begin{equation}
m^2=0\, , \quad m_{\pm}^2=-\frac{(\alpha_2+4\alpha_3)\pm \sqrt{ (\alpha_2+4\alpha_3)^2 - 4 (\beta_2+4\beta_3)}}{2(\beta_2+4\beta_3)\ell^2} \, ,
\end{equation}
which correspond, in addition to the usual massless graviton, to two new massive gravitons.

An additional simplification occurs for theories such that, besides \req{keyko}, also satisfy the condition $F_3(\ell^2 \bar\Box)=-F_2(\ell^2 \bar\Box)/4$. Those two conditions can then be rewritten as 
\begin{equation}\label{gbb}
F_1(\ell^2 \bar\Box)=F_3(\ell^2 \bar\Box)=-F_2(\ell^2 \bar\Box)/4 \, ,
\end{equation}
and, in that case, the linearized equations reduce to
\begin{equation}
\frac{\lambda}{2} G_{ab}^{(1)}=0\, ,
\end{equation}
namely, to the usual linearized Einstein equation.
Hence, for theories whose effective action satisfies the pair of conditions (\ref{gbb}), the linearized equations on Minkowski space are identical to the Einstein gravity ones ---or, in other words, the higher-derivative densities do not contribution at all to the linearized equations. Gauss-Bonnet gravity is a particular instance, which corresponds to setting all functions equal to constants, but the set of higher-derivative theories with this property contains infinitely many densities with an arbitrarily large number of covariant derivatives. We will see later that Generalized Quasi-topological gravities fall within this category (not so Brane-world gravities).


\subsubsection{Newton potential}
Here we study how the usual Newtonian potential gets modified by the introduction of higher-derivative terms as in \req{effq}. This will give us another perspective on the new types of massive modes propagated by these theories. We consider a metric perturbation on Minkowski spacetime of the form
\begin{equation}
\begin{aligned}\label{Nmetric}
    \rmd s^2_{\rm N}&=-\left[ 1+2U( r )\right]\rmd t^2+\left[1-2V( r )\right]\delta_{ij}\rmd x^i\rmd x^j\,,
\end{aligned}
\end{equation}
where $r=|\vec{x}|$ and $U( r )$ will be the Newtonian potential. Now, we evaluate (\ref{lineareqs}). We find 2 linearly independent equations for a static source in the stress tensor as $T_{00}^{(1)}=\rho(r)$,
\begin{equation}\label{eq: EOM non rel}
    \begin{aligned}
    &\mathcal{O}_{U_t} U(r)+\mathcal{O}_{V_t} V(r) =\frac{\rho(r)}{2\ell^2\lambda}~,\\
   &\mathcal{O}_{V_x} V(r)+ \mathcal{O}_{U_x}U(r)=0
\end{aligned}
\end{equation}
where
\begin{equation}
    \begin{aligned}
        \mathcal{O}_{U_t}&\equiv 2 (2F_3+2 F_1+F_2)\bar\Box ^2\, \\
        \mathcal{O}_{V_t}&\equiv(D-2)(\ell^{-2}-\bar\Box  (4 F_1+F_2))\bar\Box\, ,\\
        \mathcal{O}_{V_x}&\equiv(((-4 F_3-D (4 F_1+F_2)+8 F_1+F_2)+(D-3)\ell^{-2})\bar\Box^2\, ,\\
        \mathcal{O}_{U_x}&\equiv((4F_1+F_2)\bar\Box-\ell^{-2}))\bar\Box\, . 
    \end{aligned}
\end{equation}
We can solve this system of second-order ordinary differential equations using Fourier transforms. Denoting by $\rho_{\vec{k}}$ the Fourier transform of $\rho(r)$ in momentum space and $F_{i}^{(k)}=F_i(-\ell^2k^2)$, we find
\begin{equation}\label{eq:Integrals U V}
    \begin{aligned}
    U(r)&=\int\tfrac{\rmd^{D-1}\vec{k}}{(2\pi)^{D-1}}
   \tfrac{\rho_{\vec{k}}  \left[\vec{k}^2 \left(4F^{(\vec{k})}_3+4 F^{(\vec{k})}_1 (D-2)+(D-1)F^{(\vec{k})}_2\right)+(D-3)\ell^{-2}\right]}{4 \vec{k}^2 \left[\vec{k}^2
   \left(4F^{(\vec{k})}_3+F^{(\vec{k})}_2\right)-\ell^{-2}\right] \left[\vec{k}^2 \left(4F^{(\vec{k})}_3+4(D-1)F^{(\vec{k})}_1 +DF^{(\vec{k})}_2\right)+(D-2)\ell^{-2} \right]} \rme^{\rmi\vec{k}\cdot\vec{x}}~, \\
    V(r)&=\int\tfrac{\rmd^{D-1}\vec{k}}{(2\pi)^{D-1}}\tfrac{\rho_{\vec{k}}  \left[\vec{k}^2 \left(4 F^{(\vec{k})}_1+F^{(\vec{k})}_2\right)+\ell^{-2}\right]}{4 \vec{k}^2 \left[\vec{k}^2 \left(4F^{(\vec{k})}_3+F^{(\vec{k})}_2\right)-\ell^{-2}\right]
   \left[\vec{k}^2 \left(4F^{(\vec{k})}_3+4(D-1)F^{(\vec{k})}_1+DF^{(\vec{k})}_2\right)+(D-2)\ell^{-2}\right]}\rme^{\rmi\vec{k}\cdot\vec{x}}~.
\end{aligned}
\end{equation}
These are rather implicit formulas, but we can make further progress in the case of theories for which the functions $F_1,F_2,F_3$ are polynomials, namely,
\begin{equation}\label{eq:Nabla expansion}
    F_1=\sum_{n=0}^{N_1} {\alpha_1}_n\qty(\ell^2\bar\Box)^n,\quad F_2=\sum_{n=0}^{N_2} {\alpha_2}_n\qty(\ell^2\bar\Box)^n,\quad F_3=\sum_{n=0}^{N_3} {\alpha_3}_n\qty(\ell^2\bar\Box)^n.
\end{equation}
where ${\alpha_1}_n$, ${\alpha_2}_n$, ${\alpha_3}_n$ are constant coefficients. Introducing the notation
\begin{equation}
    N\equiv\max\qty{N_1,\,N_2,\,N_3}~,
\end{equation}
and considering a point-like source of mass $M$, we find the following result for the modified Newton potential in the most general case,\footnote{This result assumes that all the masses are different. The limit in which two or more masses coincide must be taken with care. This result also does not capture the case in which the denominators in \eqref{eq:Integrals U V} are entire functions with no zeros (besides $k^2=0$). This can only happen with an infinite number of derivatives \cite{Biswas:2011ar}.}
\begin{equation}\label{eq:U(r) Power law}
\begin{aligned}
U(r)&=-\frac{4 \Gamma[\frac{D-1}{2}]}{(D-2)  \pi^{\frac{D-3}{2}}}  \frac{G\,M}{r^{D-3}}  \left[1+r^{\frac{D-3}{2}}\sum_{i=1}^N\left[\nu_{g_{i}}K_{\frac{D-3}{2}}(m_{g_i} r)+\nu_{s_{i}}K_{\frac{D-3}{2}}(m_{s_i} r) \right]\right]\,.
\end{aligned}
\end{equation}
Here $K_{\ell}(x)$ are modified Bessel functions of the second kind, we denoted $G \equiv 1/(16\pi \lambda)$ and 
\begin{align}
    \nu_{g_{i}}\equiv-\frac{(D-2)^2m_{g_i}^{\frac{D-3}{2}}}{2^{\frac{D-1}{2}}\Gamma(\tfrac{D+1}{2})}\prod_{j\neq i}^N\qty(1-\frac{m_{g_i}^2}{m_{g_j}^2})^{-1}\, , \quad
    \nu_{s_{i}}\equiv\frac{m_{s_i}^{\frac{D-3}{2}}}{2^{\frac{D-1}{2}}\Gamma(\tfrac{D+1}{2})}\prod_{j\neq i}^N\qty(1-\frac{m_{s_i}^2}{m_{s_j}^2})^{-1}\, .
\end{align}
In these expressions, $m_{g_i}$ and $m_{s_i}$ correspond, respectively, to the masses of new spin-2 and spin-0 modes. They are nothing but the poles 
of the integrals in (\ref{eq:Integrals U V}), 
namely, the roots of
\begin{align}
m_{g_i}:\quad &k^2 (4F_3^{(k)}+F_2^{(k)})-\ell^{-2}=0~,\label{eq:m_gi}\\
m_{s_i}:\quad &k^2 (4F_3^{(k)}+4 (D-1)F_1^{(k)} +DF_2^{(k)})+(D-2)\ell^{-2}=0~.\label{eq:m_si}
\end{align}
{The net negative contributions in the Newtonian potential indicate which of the modes are ghosts. In the case of constant $F_i$ previously studied in \cite{Aspects,Sisman:2011gz}, the scalar mode is always contributing positively and is never a ghost, whereas the opposite holds for the massive spin-2 mode, which is always a ghost. In the general case, we observe that some of the scalar modes can also be ghosts, while some of spin-2 modes can carry positive energy.
Note also that for theories satisfying condition (\ref{keyko}), the second equation becomes rootless and there are no new scalars, in agreement with the analysis of the previous subsection. On the other hand, if $F_3(\ell^2 \bar \Box)=-F_2(\ell^2 \bar \Box)/4$ holds then there are no new spin-2 modes. If both conditions hold at the same time, the Newton potential reduces to the Einstein gravity one.

On the other hand, we find for the other metric function,
\begin{equation}\label{eq:V eff}
\begin{aligned}
V(r)&=-\frac{2\Gamma[\frac{D-3}{2}]}{(D-2)  \pi^{\frac{D-3}{2}}}  \frac{G\,M}{r^{D-3}}\left[1+r^{\frac{D-3}{2}}\sum_{i=1}^N\left[\tilde{\nu}_{g_{i}}K_{\frac{D-3}{2}}(m_{g_i} r)+\tilde{\nu}_{s_{i}}K_{\frac{D-3}{2}}(m_{s_i} r) \right]\right]\,
\end{aligned}
\end{equation}
where
\begin{align}
\tilde{\nu}_{g_{i}}&\equiv\frac{- (D-2)m_{g_i}^{\frac{D-3}{2}}}{2^{\frac{D-5}{2}}(D-1)\Gamma(\tfrac{D-3}{2})}\prod_{j\neq i}^N\qty(1-\frac{m_{g_i}^2}{m_{g_j}^2})^{-1}\, , \quad
\tilde{\nu}_{s_{i}}&\equiv\frac{-m_{s_i}^{\frac{D-3}{2}}}{2^{\frac{D-5}{2}}(D-1)\Gamma(\tfrac{D-3}{2})}\prod_{j\neq i}^N\qty(1-\frac{m_{s_i}^2}{m_{s_j}^2})^{-1}\, .
\end{align}
As expected, the expressions above reduce to the results in \cite{Aspects} for the potentials $U(r)$ and $V(r)$ when $N=0$ in (\ref{eq:U(r) Power law}, \ref{eq:V eff}).

\section{Generalized Quasi-topological gravities in $D=4$}\label{gqT}

In this section we present the first examples of Generalized Quasi-topological (GQT) densities involving covariant derivatives of the Riemann tensor. We focus on $D=4$. In that number of dimensions, in the absence of covariant derivatives it has been shown that there exists a unique non-trivial GQT density at each curvature order. Here we show that the landscape of GQT theories is modified considerably by allowing covariant derivatives of the Riemann tensor to appear in the action. In particular, while we find no new densities at four- and six-derivative(s of the metric) orders, we obtain four new inequivalent GQTs at eight-derivative order. Of these, only one possesses an integrated equation for $f(r)$ which is of second order in derivatives, two of them have third-order equations, and the remaining one has an integrated fourth-order equation for the metric function. In all cases, we find that the Schwarzschild solution is also a solution of these theories. As a consequence, coupling Einstein gravity to these theories does not give rise to new spherically symmetric black hole solutions. Extending the analysis to ten-derivative order, we find new examples which do not  admit  Schwarzschild as a solution. For those, the coupling to Einstein gravity does produce new non-trivial modifications of the Schwarzschild black hole. Similarly to what happens for polynomial GQTs, we find that the thermodynamic properties of those solutions can be computed analytically. We study the relation between their temperature and their mass and find a deviation from the universal behaviour previously observed in the case of general polynomial GQTs for small black holes. Instead of the prototypical $T\sim M^{1/3}$ scaling universally found for such theories \cite{PabloPablo4, Bueno:2019ycr}, the density with covariant derivatives induces a different behaviour of the form $T\sim M^3$.


Let us start by recalling the basic definition and properties of GQTs. 
Consider a general static and spherically symmetric (SSS) spacetime parametrized by two functions, 
$N(r)$ and $f(r)$,
\begin{equation}
\label{SSS}
\diff s^2_{N, f}=-  N(r)^2 f(r) \diff t^2+\frac{\diff  r^2}{f(r)}  + r^2\diff \Omega^2_{(D-2)}\, ,
\end{equation}
where $\diff \Omega^2_{(D-2)}$ is the  $(D-2)$-dimensional sphere metric. The following comments extend, with minor modifications, to the cases in which the horizon is  hyperbolic or planar instead. The expressions below will incorporate those cases through a parameter  denoted $k$ which will take the values $+1,0,-1$, respectively for the spherical, planar and hyperbolic cases.

For a given curvature invariant of order $2m$ in derivatives of the metric and involving $p$ covariant derivatives of the Riemann tensor, $\mathcal{R}_{(2m,p)}$, let  $S_{N,f}$ and $L_{N,f} $ be, respectively, the  effective on-shell action and Lagrangian resulting from the evaluation of  $\sqrt{|g|}\mathcal{R}_{(2m,p)}$ in the ansatz (\ref{SSS}), namely,
\begin{equation}\label{ansS}
L_{N,f}\equiv \left. N(r)   r^{D-2}   \mathcal{R}_{(2m,p)}\right|_{N,f}\, ,   \quad  S_{N,f}\equiv   \Omega_{(D-2)}\int \diff t  \int  \diff r L_{N,f} \, ,
\end{equation}
where we performed the trivial integral over the angular directions,   $\Omega_{(D-2)}\equiv   2\pi^{\frac{D-1}{2}}/\Gamma[\frac{D-1}{2}]$. We denote by $L_f\equiv L_{1,f}$ and $S_f\equiv S_{1,f}$ the expressions resulting from setting $N=1$  in $L_{N,f}$. 
Now, solving the full nonlinear equations of motion for a metric of the form (\ref{SSS}) can be shown to be equivalent to solving  the Euler-Lagrange equations of $S_{N,f}$ associated to $N(r)$ and $f(r)$ \cite{Palais:1979rca,Fels:2001rv,Deser:2003up,PabloPablo4}, namely,
\be
\left.\mathcal{E}^{ab}\right|_{N,f}\equiv \left. \frac{1}{\sqrt{|g|}}  \frac{\delta S}{\delta g^{ab}}  \right|_{N,f}=0 \quad   \Leftrightarrow   \quad \frac{\delta S_{N,f}}{\delta N}= \frac{\delta S_{N,f}}{\delta f}=0\, .
\ee
We say that  $\mathcal{R}_{(2m,p)}$ is a GQT density if the Euler-Lagrange equation of $f(r)$ associated to  $S_f$ is identically   vanishing, namely,  if
\begin{equation} \label{GQTGcond}
\frac{\delta S_{f}}{\delta f}=0\, ,   \quad \forall \, \, f(r)\, .
\end{equation}
This condition is equivalent to  asking $L_f$ to be  a total derivative,  
\begin{equation}\label{condd2}
L_f =T_0'\, ,
\end{equation}
for certain function  $T_0(r,f(r),f'(r),\dots , f^{(p+1)})$.

Thus, the variation with respect to $f(r)$ of the on-shell action $S_f$ determines whether or not a given density is of the  GQT class. When that is the case, the full non-linear equations of $\mathcal{R}_{(2m,p)}$ reduce to a single equation for $f(r)$ which can in fact be integrated once. Such integrated equation can be obtained from the variation of $L_{N,f}$ with  respect to $N(r)$ as \begin{equation}\label{eqf}
\left.\frac{\delta S_{N,f}}{\delta N}\right|_{N=1}=0\, \quad \Leftrightarrow   \quad   \text{equation of}\quad f(r)\, .
\end{equation}
Let us see this in more detail. As explained in \cite{PabloPablo3}, whenever \req{condd2} holds, the effective Lagrangian  $L_{N,f}$ takes the form
\begin{equation}\label{fofwo}
L_{N,f}=N T_0' +  N' T_1 +  N'' T_2+ \dots + N^{(p+2)}T_{p+2} +\mathcal{O}(N'^2/N)\, ,
\end{equation}
where $T_{1},T_2,\dots,T_{p+2}$  are functions of $f(r)$ and its derivatives (up to $f^{(p+2)}$), and $\mathcal{O}(N'^2/N)$ is a sum of  contributions  which are all at least quadratic in  derivatives of   $N(r)$. Integrating by parts one finds
\begin{equation}
S_{N,f} = \Omega_{(D-2)}  \int \diff t \int \diff r \left[N\left(T_0 + \sum_{j=1}^{p+2} (-1)^{j} T_{j}^ {(j-1)}\right)' +\mathcal{O}(N'^2/N) \right]\, .
\end{equation}
Therefore, one can write every term involving one power of $N(r)$  or  its derivatives as certain  product of $N(r)$  and a total derivative which depends on $f(r)$ alone. As a consequence, eq.~(\ref{eqf}) equates such a total derivative to zero. Integrating it once one we are left with \cite{PabloPablo3}
\begin{equation} \label{eqqqf}
\mathcal{F}_{\mathcal{R}_{(2m,p)}}  \equiv  T_0 + \sum_{j=1}^{p+2} (-1)^{j} T_{j}^ {(j-1)}=\frac{M}{\Omega_{(D-2)}}\, ,
\end{equation}
where the integration constant was written in terms of the ADM mass of the solution \cite{Arnowitt:1960es,Arnowitt:1960zzc,Arnowitt:1961zz,Deser:2002jk}.

In sum, given some linear combination of GQT densities, the equation satisfied by $f(r)$ can be obtained from  $L_{N,f}$ as defined in \req{ansS} by identifying the functions $T_{\{ j \}}$ from \req{fofwo}. The order of the integrated equation $\mathcal{F}_{\mathcal{R}_{(2m,p)}} $ is at least two orders less than the one of the equations determining $f(r)$ and $N(r)$ in the most general case, namely, 
\begin{equation}
\mathcal{F}_{\mathcal{R}_{(2m,p)}}=\mathcal{F}_{\mathcal{R}_{(2m,p)}}(r,f,f',\dots,f^{(2p+2)})
\end{equation}
In particular, when $p=0$, corresponding to the case without covariant derivatives of the Riemann tensor, the integrated equation is at most second-order in derivatives of $f(r)$. In that case, one can see that the integrated equations are either of order $0$ in derivatives ---these are called simply ``Quasi-topological'' theories \cite{Quasi2,Quasi,Dehghani:2011vu,Ahmed:2017jod,Cisterna:2017umf}, which includes Lovelock theories \cite{Lovelock1,Lovelock2} as particular cases--- or, alternatively, of order $2$.  As we will see in a moment, the actual order of the integrated equations that we will find in our new GQT densities with covariant derivatives will be considerably lower than the $2p+2$ upper bound.

 
We will say that two GQT densities $\{ \mathcal{R}^{I}_{(2m,p)} ,\mathcal{R}^{II}_{(2m,p)} \}$ are ``inequivalent'' (as far as  SSS  solutions are  concerned) whenever the quotient of their respective integrated equations is not constant, namely,
\begin{equation}
\mathcal{R}^{I}_{(2m,p)} \quad \text{inequivalent from} \quad \mathcal{R}^{II}_{(2m,p)} \quad \Leftrightarrow \quad  \frac{\mathcal{F}_{\mathcal{R}^I_{(2m,p)}}(r,f,f',\dots,f^{(2p+2)}) }{\mathcal{F}_{\mathcal{R}^{II}_{(2m,p)}} (r,f,f',\dots,f^{(2p+2)})} \neq \text{constant} \, .
\end{equation}
Otherwise we will call them ``equivalent''. Two equivalent densities differ by densities which make no contribution whatsoever to the integrated equation of $f(r)$. Those densities are ``trivial'' as far as SSS solutions are concerned. 

 In the $p=0$ case, it has been argued that: i) there exist no (non-trivial) GQTs in $D=3$ \cite{Bueno:2022lhf}; ii) there exists a single inequivalent GQT density at each curvature order $m$ in $D=4$ whose integrated equation is a differential equation of order $2$ \cite{Moreno:2023rfl}; there exists a single inequivalent Quasi-topological density at each curvature order $m$ in $D\ge 5$ whose integrated equation is algebraic  \cite{Bueno:2022res}; there exist $(m-2)$ inequivalent GQT densities at each  curvature order in $D\ge 5$ whose integrated equation is a differential equation of order $2$  \cite{Bueno:2022res,Moreno:2023rfl}.



\subsection{Linear spectrum}\label{subsec:linearGQT}
A remarkable property of all GQTs built from polynomial curvature invariants is that their linear spectrum on maximally symmetric backgrounds is devoid of ghosts. In fact, the linearized equations of motion are proportional to those of Einstein gravity on the same background. In the case of polynomial GQTs, the second-order nature of the linearized equations was first verified explicitly in case-by-case examples ---see \eg \cite{Quasi2,Quasi, PabloPablo, Hennigar:2017ego, Ahmed:2017jod}. It was subsequently proven that the single-metric-function condition that defines GQTs also implies the linearization is second-order in general~\cite{PabloPablo3} ---c.f. page 102 of \cite{CanoMolina-Ninirola:2019uzm} for the most up-to-date version of this proof. 
Here we show that this result in fact holds for all GQTs, including those that contain covariant derivatives of the curvature (and hence have equations of motion of order greater than four). 

The idea behind the proof consists in considering a metric perturbation within the single-function static spherically symmetric ansatz. Thus, we start by considering the metric \eqref{SSS} with $N(r)=1$. For convenience, let us rewrite this metric as 
\begin{equation}
ds^2=-f(r)\diff u^2-2 \diff r \diff u+r^2\diff\Omega_{(D-2)}^2\, ,
\end{equation}
where $u=t+r_{*}$, and where $r_*$ is the tortoise coordinate, defined by $\diff r_{*}=\diff r/f(r)$. One can show that in this coordinate system the GQT condition \eqref{GQTGcond} is equivalent to the vanishing of the $rr$ component of the equations of motion, that is, 
\begin{equation}\label{GQTGcond2}
\mathcal{E}_{rr}=0\, ,\quad \forall f(r)\, .
\end{equation}
We then take $f(r)$ to be 
\begin{equation}\label{radialpert}
f(r)=1+\frac{r^2}{\ell_{\star}^2}+h(r)\, ,\quad h(r)\ll 1\, ,
\end{equation}
corresponding to a maximally symmetric vacuum plus a small perturbation $h_{ab}$ given by
\begin{equation}\label{radialperthab}
h_{uu}=h(r)\, .
\end{equation} 
Then, the idea is to impose the condition \eqref{GQTGcond2} at the level of the linearized equations by using this perturbation. We know that, in general, the linearized equations are given by \eqref{eq:Eab general msb} for certain coefficients $\alpha_l$, $\beta_l$ and $\gamma_l$. Let us for instance assume that our theory has sixth-order equations of motion --- so that only the coefficients with $l\le 2$ are nonzero --- and let us set $D=4$. We get, after a direct evaluation of \eqref{eq:Eab general msb} on \eqref{radialperthab},
\begin{equation}\label{Errlinear}
\begin{aligned}
\mathcal{E}_{rr}^{(1)}&=\alpha_1 \ell^2\left(-\frac{4 h}{r^4}+\frac{2 h''}{r^2}\right)+\gamma _1\ell^2 \left(-\frac{12 h}{r^4}+\frac{6 h''}{r^2}-\frac{4 h^{(3)}}{r}-h^{(4)}\right)\\
&+\alpha _2 \ell^4 \left(h \left(-\frac{32}{r^6}+\frac{56}{L^2 r^4}\right)+\frac{32 h'}{r^5}+\left(-\frac{16}{r^4}-\frac{28}{L^2 r^2}\right) h''+\frac{16 h^{(3)}}{L^2
   r}+\left(\frac{4}{L^2}+\frac{4}{r^2}\right) h^{(4)}\right)\\
   &+\gamma _2 \ell^4\left(h \left(-\frac{80}{r^6}+\frac{120}{L^2 r^4}\right)+\frac{80
   h'}{r^5}+\left(-\frac{40}{r^4}-\frac{60}{L^2 r^2}\right) h''+\frac{40 h^{(3)}}{L^2 r}+\left(-\frac{20}{L^2}+\frac{10}{r^2}\right) h^{(4)}\right.\\
&\left.+\left(-\frac{6}{r}-\frac{12
   r}{L^2}\right) h^{(5)}+\left(-1-\frac{r^2}{L^2}\right) h^{(6)}\right)\, .
   \end{aligned}
\end{equation}
Then, the GQT condition \eqref{GQTGcond2} implies that this must vanish for any choice of $h(r)$. Clearly, this only happens if $\alpha_1=\alpha_2=\gamma_1=\gamma_2=0$, since all the terms are linearly independent. The same conclusion follows in general dimensions and if the theory has higher-order equations of motion. In the latter case \req{Errlinear} will include $\alpha_l$- and $\gamma_l$-terms with higher $l$, but these are all linearly independent because they contain different numbers of derivatives of $h$ and/or different radial dependence. 

In conclusion, \eqref{GQTGcond2} implies the vanishing of all the $\alpha_{l}$ and $\gamma_{l}$ except for $\alpha_0$, corresponding to the coefficient of the linearized Einstein tensor. Finally, the relation \eqref{f2relation} implies the vanishing of the $\beta_{l}$ coefficients. Therefore, the linearized equations must be proportional to the linearized Einstein tensor.

\subsection{Classification of four-dimensional theories}
\label{sec:4dTheories}

In this section, we will classify all possible GQT Lagrangians, based on the number of derivatives of the metric appearing in the action. In the case of four and six derivatives, the result is in line with previous considerations~\cite{Quasi2, Quasi, Hennigar:2017ego, PabloPablo3}: nothing new beyond those theories constructed from the polynomial invariants is found. However, the cases of eight and ten derivatives reveals new features not seen before. 

Let us briefly summarize the methodology. At a given derivative order, we construct the most general Lagrangian density by performing a linear combination of all curvature invariants that appear at that order:
\be 
\mathcal{L}_{(2m)} = \sum_{i} c_{(i)}^{(2m,p)} \mathcal{R}^{(i)}_{(2m,p)} \, .
\ee
Here, $2m$ refers to the number of derivatives of the metric appearing in the term, while the $c_{(i)}^{(2m,p)}$'s are constants. The densities $ \mathcal{R}^{(i)}_{(2m,p)} $ involve contractions of the Riemann tensor and its covariant derivatives. In appendix \ref{basis} we present a generating set of these invariants for up to eight derivatives of the metric. The action is then evaluated on a single-function SSS metric ansatz and we impose \req{GQTGcond}, namely, that the Euler-Lagrange equation for $f(r)$ vanishes. This leads to constraints on the $c_{(i)}^{(2m,p)}$'s such that the resulting theory is of the GQT  type. 

Let us make a few further comments regarding the densities involving derivatives of the curvature. In general it is possible to reduce the number of invariants that make non-trivial contributions to the equations of motion by integrating by parts and utilizing the Bianchi identities. However, we have not pursued this option here. The reasons are simply because, at high-order in derivatives, there are so many terms that it would be impractical to do so. Furthermore, as will be obvious below, it is not necessary to do this  to understand the effects of these terms. Therefore, in constructing our actions at the four, six, and eight-derivative levels, we include all possible terms at a given order (as listed in the appendix). On the other hand, in the case of ten-derivative theories our analysis will not be exhaustive. 

\subsubsection{Two-derivative actions}

For completeness, we include here the two-derivative sector, which is simply Einstein gravity, 
\begin{equation}
\mathcal{L}_{(2,0)}^{(1)}= R \, .
\end{equation}
The integrated equation for the metric function is given by
\be \label{EEEin}
\mathcal{F}^{(1)}_{(2,0)} = -2 r (f-k) \, .
\ee

\subsubsection{Four-derivative actions}

There are no non-trivial four-derivative GQT actions in four-dimensions.

\subsubsection{Six-derivative actions}

There is a single non-trivial six-derivative GQT action in four-dimensions. The action for this theory may be taken to be that of Einsteinian Cubic Gravity  \cite{PabloPablo}
\begin{equation}
\mathcal{L}_{(6,0)}^{(1)}=+12 \tensor{R}{_{a}^{c}_{b}^{d}}\tensor{R}{_{c}^{e}_{d}^{f}}\tensor{R}{_{e}^{a}_{f}^{b}}+R_{ab}^{cd}R_{cd}^{ef}R_{ef}^{ab}-12R_{abcd}R^{ac}R^{bd}+8R_{a}^{b}R_{b}^{c}R_{c}^{a}\, ,
\end{equation}
whose integrated equation for the metric function $f(r)$ reads \cite{Hennigar:2016gkm,PabloPablo2}
\begin{align}
\mathcal{F}_{(6,0)}^{(1)} &= - \frac{4}{r^2} \bigg[ r f f'' \left(rf' + 2(k-f) \right) - \frac{f'}{3} \left(r^2 f'^2 + 3 r k f' + 6 f(k-f) \right)\bigg] \, .
\end{align}

\subsubsection{Eight-derivative actions}
There are five non-trivial eight-derivative GQTG actions in four-dimensions. The first of these possibilities may be taken to be that given by the standard polynomial invariants ---see \eg \cite{PabloPablo4}. However, the additional four theories require terms involving covariant derivatives of the Riemann tensor. Of these, a single combination can be formed such that the integrated equations are second-order, while the remaining three involve higher-derivatives of the metric function. As examples of actions that give rise to each of the new sets of GQTGs, the following choices may be made:
\begin{align}
\mathcal{L}_{(8,0)}^{(1)} =&+ \tensor{R}{^{pqrs}} \tensor{R}{_p^t_r^u} \tensor{R}{_t^v_q^w} \tensor{R}{_{uvsw}}  - \frac{13}{5} \tensor{R}{^{pqrs}} \tensor{R}{_{pq}^{tu}} \tensor{R}{_r^v_t^w} \tensor{R}{_{svuw}}  - \frac{1}{8} \tensor{R}{^{pqrs}} \tensor{R}{_{pq}^{tu}} \tensor{R}{_{tu}^{vw}} \tensor{R}{_{rsvw}}  
\nonumber\\
&+ \frac{1}{5} R \tensor{R}{^{pqrs}} \tensor{R}{_q^t_s^u}\tensor{R}{_{tpur}}  \, ,
\\
\mathcal{L}_{(8,2)}^{(2)} =&+ \tensor{R}{^{pqrs}} \tensor{R}{^t_p^u_r ^{;v}} \tensor{R}{_{tqus;v}} - \tensor{R}{^{pq;r}} \tensor{R}{^{st}_p^u} \tensor{R}{_{stqr;u}} + 2 \tensor{R}{^{pqrs}} \tensor{R}{_p ^{tuv}} \tensor{R}{_{qtru;sv}} + \tensor{R}{^{pq}} \tensor{R}{^{rs;t}} \tensor{R}{_{rtsp;q}} 
\nonumber\\
&- 2 \tensor{R}{^{pq;r}} \tensor{R}{_p ^{s;t}} \tensor{R}{_{qsrt}} + \tensor{R}{^{pq;rs}} \tensor{R}{^t_p^u_r}\tensor{R}{_{tqus}} 
-\tensor{R}{^{pq}} \tensor{R}{^{rs}_{;p}} \tensor{R}{_{rs;q}} - \frac{1}{2} \tensor{R}{^{;p}} \tensor{R}{^{qr;s}} \tensor{R}{_{pqrs}}\nonumber\\
& + \tensor{R}{^{pq}} \tensor{R}{^{rs}_{;q}^t} \tensor{R}{_{prst}} 
- \frac{1}{2} \Box  \tensor{R}{^{pq}} \tensor{R}{^{rst}_p} \tensor{R}{_{rstq}} 
\\
\mathcal{L}_{(8,4)}^{(3)} =&+ 21 R^{pq;rst} R_{prqs;t} - 12 \Box R^{pq} \Box R_{pq} - 12 R^{pq} R^{rs} R_{pq;rs} + 153 R^{pq} R^{rs;t} R_{rtsp;q} + 6 R^{;pq} \Box R_{pq} 
\nonumber\\
&+ \frac{33}{4} R^{pqrs;tu} R_{pqrs;tu} + 6 R^{;pq} \tensor{R}{^{rst}_q} R_{rstp} - \frac{21}{2} \Box R^{pq} \tensor{R}{^{rst}_p} R_{rstq} + 36 R^{pq;rs} \tensor{R}{^t_p^u_r} R_{tqus} 
\nonumber\\
&+ 183 R^{pqrs} \tensor{R}{_p^{tuv}} R_{qtru;sv} - 51 R^{pq} \tensor{R}{^{rs}_{;p}} R_{rs;q} - \frac{177}{2} R^{;p} R^{qr;s} R_{pqrs} - 81 R^{pq;r} \tensor{R}{_p^{s;t}} R_{qrst} 
\nonumber\\
&- 93 R^{pq;r} \tensor{R}{_p^{s;t}} R_{qsrt} - 60 R^{pq;r} \tensor{R}{^{st}_p^u} R_{stqr;u} + 33 R^{pqrs} \tensor{R}{^{tuv}_{p;q}} R_{tuvr;s} \nonumber\\
& - 27 R^{pqrs} \tensor{R}{^{tuv}_{p;r}} R_{tuvq;s} 
- 63 R^{pqrs} \tensor{R}{^t_p^u_r^{;v}} R_{tqus;v}
\\
\mathcal{L}_{(8,4)}^{(4)} =&+ 52 R^{p q} R^{rs;t} R_{rtsp;q} + 8 R^{pq;rst} R_{prqs;t}  -4 \Box R^{pq} \Box R_{pq} -20 R^{pq;r} \tensor{R}{_p^{s;t}} R_{qsrt} \nonumber\\
&- 24 R^{pq;r} \tensor{R}{^{st}_p^u} R_{stqr;u} 
-4 R^{pq;r} \tensor{R}{^s_p^{tu}} R_{sqtr;u} + 12 R^{pqrs}\tensor{R}{^{tuv}_{p;q}} R_{tuvr;s} \nonumber\\
& - 10 R^{pqrs} \tensor{R}{^{tuv}_{p;r}} R_{tuvq;s} -20 R^{pqrs} \tensor{R}{^t_p^u_r^{;v}} R_{tqus;v} 
+ 2 R^{;pq} \Box R_{pq} + 3R^{pqrs;tu} R_{pqrs;tu} \nonumber\\
&+ 72 R^{pqrs} \tensor{R}{_p^{tuv}} R_{qtru;sv} - 8R^{pq} \tensor{R}{^{rs}_{;p}} R_{rs;q} -22 R^{;p} R^{qr;s} R_{pqrs} - 36 R^{pq;r} \tensor{R}{_p^{s;t}} R_{qrst}
\\
\mathcal{L}_{(8,4)}^{(5)} =&+ 1178 R^{pq} R^{rs;t} R_{rtsp;q} + 171 R^{pq;rst} R_{prqs;t} -95 \Box R^{pq} \Box R_{pq} -76 R^{pq} R^{rs} R_{pq;rs} \nonumber\\
&- 646 R^{pq;r} \tensor{R}{_p^{s;t}} R_{qsrt} 
-475 R^{pq;r} \tensor{R}{^{st}_p^u} R_{stqr;u} + 228 R^{pq;r} \tensor{R}{^s_p^{tu}} R_{sqtr;u} \nonumber\\
&+ 266 R^{pqrs} \tensor{R}{^{tuv}_{p;q}} R_{tuvr;s} - 209 R^{pqrs} \tensor{R}{^{tuv}_{p;r}} R_{tuvq;s} 
-494 R^{pqrs} \tensor{R}{^t_p^u_r^{;v}} R_{tqus;v} \nonumber\\
&+ \frac{95}{2} R^{;pq} \Box R_{pq} + \frac{133}{2} R^{pqrs;tu}R_{pqrs;tu} + 38 R^{;pq} \tensor{R}{^{rst}_q} R_{rstp} 
+ 228  R^{pq;rs} \tensor{R}{^t_p^u_r} R_{tqus}\nonumber\\
& + 1520 R^{pqrs} \tensor{R}{_p^{tuv}} R_{qtru;sv} - 342 R^{pq} \tensor{R}{^{rs}_{;p}} R_{rs;q} - 646  R^{;p} R^{qr;s} R_{pqrs}
 \nonumber\\
&- 646 R^{pq;r} \tensor{R}{_p^{s;t}} R_{qrst}
\end{align}
The integrated equations for each of these densities read, respectively,
\begin{align}
\mathcal{F}_{(8,0)}^{(1)} &= - \frac{12 f'}{5r^3} \bigg[ \frac{r f f''}{2} \left(r f' + 2 (k-f) \right) - \frac{f'}{3} \left(\frac{3 r^2 f'^2}{8} +  \frac{r f'}{2} (f + 2k) + 3 f (k-f) \right) \bigg]\, ,
\\
\mathcal{F}_{(8,2)}^{(2)} &= -\frac{4 f^2}{r^5} \alpha^2 \, ,
\\
\mathcal{F}_{(8,4)}^{(3)} &=+ \frac{3 f^2}{2 r^5} (5 \alpha^2-2r  \alpha \alpha'+r^2 \alpha'^2) \, ,
\\
\mathcal{F}_{(8,4)}^{(4)} &= - \frac{2 f^2}{r^5} \alpha (4 \alpha-\alpha'' r^2) \, ,
\\
\mathcal{F}_{(8,4)}^{(5)} &= +\frac{19 f^2}{4r^5} (4(\alpha-\alpha' r )\beta+\alpha'' r^2 (\alpha+\beta))\,.
\end{align}
where we defined the functions\footnote{The functions $\alpha(r)$ and $\beta(r)$ are directly proportional to the non-trivial components of the traceless Ricci tensor and Weyl tensor for the single-function static, spherically symmetric background, respectively.}
\begin{equation}
\alpha(r)\equiv 2(k-f(r))+r^2 f''(r)\, , \quad \beta(r)\equiv 2(k-f(r))+2rf'(r)-r^2 f''(r)\, .
\end{equation}
From the densities involving covariant derivatives, while the first three exclusively depend on $\alpha(r)$ and its derivatives, the fourth one also includes a dependence on $\beta(r)$ ---which can not be expressed in terms of $\alpha(r)$ and its derivatives.

Observe that $\alpha(r)$ and $\beta(r)$ identically vanish when evaluated for a maximally symmetric background. Namely, if we set 
\begin{equation}\label{ads}
f(r)|_{\rm (A)dS}\equiv \frac{r^2}{L_{\star}^2}+k \quad  \Rightarrow \quad \alpha(r)|_{\rm (A)dS}=\beta(r)|_{\rm (A)dS}=0\, ,
\end{equation}
 and therefore
\begin{equation}
\mathcal{F}_{(8,2)}^{(2)}|_{\rm (A)dS}=\mathcal{F}_{(8,4)}^{(3,4,5)}|_{\rm (A)dS}=0\, ,
\end{equation}
or, in other words, the equations of motion of the new GQTs identically vanish for maximally symmetric backgrounds. Furthermore, it is easy to see that the usual Schwarzschild-(A)dS solution satisfies the equations of the new densities. This follows from the fact that 
\begin{equation}
\alpha(r)|_{ \text{Sch-(A)dS}}=0\, , \quad \beta(r)|_{ \text{Sch-(A)dS}}=\frac{12M}{r}\, ,
\end{equation}
where
\begin{equation}
f(r)|_{\text{Sch-(A)dS}}\equiv \frac{r^2}{L_{\star}^2}+k-\frac{2M}{r}\, .
\end{equation}
Since all terms appearing in $\mathcal{F}_{8}^{(i=2,3,4,5)}$ are proportional to $\alpha(r)$ or its derivatives, it follows that 
\begin{equation}
\mathcal{F}_{(8,2)}^{(2)}|_{\rm Sch-(A)dS}=\mathcal{F}_{(8,4)}^{(3,4,5)}|_{\rm Sch-(A)dS}=0\, .
\end{equation}
This implies that if we couple the new densities to Einstein gravity, the Schwarzschild solution will not receive corrections from such terms. As we explore in Appendix \ref{8dpure}, new solutions do exist when the new densities are considered as full theories by themselves, but these are less interesting. In order to obtain GQTs which give rise to continuous modifications of the Einstein gravity Schwarzschild solution we need to move up yet another curvature order.

\subsubsection{Ten-derivative actions}

To the best of our knowledge, a full classification of curvature invariants at ten-derivative order has not been undertaken. Therefore, our analysis in this section is necessarily incomplete but, as we shall see, interesting.

To study ten-derivative actions we do the following. We construct all possible combinations of ten-derivative actions built from lower-order densities ---for example, by multiplying all six-derivatives densities by the four-derivative ones, and so on. In addition to this, we include 20 additional terms that are explicitly order ten in derivatives. We list the ones used for this purpose in appendix \ref{basis}. However, particularly relevant is the following density,
\be 
C_{abcd} C^{abcd} C_{ef rs ;u} C^{ef rs ;u} \, .
\ee
As discussed in~\cite{Ruhdorfer:2019qmk,Li:2023wdz}, in four space-time dimensions there are four non-trivial parity-preserving contributions to the effective field theory of gravity at the ten-derivative level. Two of them involve the square of a dual Riemann tensor and hence they vanish identically on spherically symmetric spacetimes. We thus are left with two contributions that modify spherically symmetric solutions. The first contribution can be taken, as usual, to be a contraction of five Weyl tensors. The density appearing above is a particular choice for the second non-trivial contribution. 

The ten-derivative action is the first instance where more than one non-trivial contribution to the EFT appears. Moreover, it is the first instance where terms involving covariant derivatives of the metric play an essential role ---\ie cannot be removed by field redefinitions. For these reasons, we expected to find novel GQT theories at this order that explicitly modify the solutions to vacuum Einstein gravity, corresponding to the two possible non-trivial effective field theory contributions. This expectation will be borne out.  

From the entire set of ten-derivative invariants that we construct, there turn out to be 21 independent contributions. This represents notable growth compared to the eight-derivative case where there were five independent contributions. Of the 21 independent ten-derivative GQT theories, only two of these are non-trivial when evaluated on the Schwarzschild solution ---corresponding to $\mathcal{F}_{(10, 0)}^{(1)}$ and $\mathcal{F}_{(10, 4)}^{(9)}$ below. Of the 21 theories, 5 have second-order integrated equations, 7 have third-order, 6 have forth-order, 2 have fifth-order, and 1 has sixth-order. As we have not included all possible 10 derivative densities in our starting action, these numbers are likely to be incomplete. However, we expect that any additional GQTs, should they exist, will not correct the solutions of vacuum general relativity. The list of 21 inequivalent integrated equations reads
\begin{align}
\mathcal{F}_{(10, 0)}^{(1)} =&+ \frac{f'^2}{r^2} \bigg[\frac{f'^3}{5} + \frac{2(f+k) f'^2}{4 r} - \frac{2f(f-k)f'}{r^2} - \frac{f f''}{r} \left( r f' + 2(k-f) \right) \bigg] \, ,
\\
\mathcal{F}_{(10, 4)}^{(2)} =&+ \frac{f^2 \alpha^2 (f-k)}{r^7} \, , 
\\
\mathcal{F}_{(10, 4)}^{(3)} =&+ \frac{f^2 \alpha (\alpha + \beta)^2}{r^7} \, ,
\\
\mathcal{F}_{(10, 4)}^{(4)} =&+ \frac{f^2 \alpha \left( 8 \alpha (f-k)  + \alpha^2 - \beta^2 \right)}{r^7} \, ,
\\
\mathcal{F}_{(10, 4)}^{(5)} =&+ \frac{f^2 \alpha^2 \left(6(k-f) - \alpha \right)}{r^7} \, ,
\\
\mathcal{F}_{(10, 4)}^{(6)} =&+ \frac{f^2 (\alpha + \beta)^2 \left(r \alpha' - 2 \alpha  \right)}{r^7} 
\\ 
\mathcal{F}_{(10, 4)}^{(7)} =&+ \frac{f^2 \alpha (\alpha + \beta) \left( r \alpha' - \alpha + \beta \right)}{r^7} \, ,
\\
\mathcal{F}_{(10, 4)}^{(8)} =& -\frac{f^2 \alpha \beta \left(r \alpha' - \alpha + \beta \right)}{r^7}  \, , 
\\
\mathcal{F}_{(10, 4)}^{(9)} =&+ \frac{1}{r^7} \bigg[ 48 f^2 r^3 \left(-r^2 f'' + 2 r f' + 2 (k-f) \right) (k-f) f''' 
\nonumber
\\
&+  f \left(2 \left( 65 f + 16 k \right) r f' 
+ 4 \left(2 k - 65 f \right) (k-f) \right) \left(\frac{r f'}{2} + k - f \right) r^2 f'' - 4 k r^4 f'^4 
\nonumber
\\
&- 3 \left(3 k^2 + 4 k f + 121 f^2 \right) r^3 f'^3 - 2 \left(2 k^2 + 38 k f + 1271 f^2 \right)(k-f) r^2 f'^2 
\nonumber
\\
&- 40 r f (k + 122 f) \left(k-f \right)^2 f' - 3448 f^2 (k-f)^3 \bigg] \, ,
\\
\mathcal{F}_{(10, 4)}^{(10)} =&+ \frac{f^2}{r^7} \bigg[4 r^2 (k-f) \alpha'^2 - 2 r \left( 4(k-f) + \alpha + \beta \right) \alpha \alpha'  - \alpha (\alpha + \beta)(\beta - 3 \alpha) \bigg] \, , 
\\
\mathcal{F}_{(10, 4)}^{(11)} =&+ \frac{f^2}{r^7} \bigg[ r^2 \left(4(k-f) - \alpha -\beta \right) \alpha'^2 - 2 r \left(\left(4(k-f) + \beta \right) \alpha + \beta^2 \right) \alpha'  + 3 \alpha (\alpha + \beta)^2 \bigg] \, , 
\\
\mathcal{F}_{(10, 4)}^{(12)} =&+\frac{f^2}{r^7} \bigg[2r^2 \left(2(k-f) - \alpha \right) \alpha'^2 - 2 r \left(4(k-f) + 3 \beta - \alpha \right) \alpha \alpha' + \alpha \left(\alpha^2  + 6 \alpha \beta  -3 \beta^2 \right) \bigg] \, , 
\\
\mathcal{F}_{(10, 4)}^{(13)} =&+ \frac{f^2}{r^7} \bigg[ r^2 (k-f) (\alpha +\beta) \alpha'' - r \left(\alpha^2 + \alpha \beta + 4 \beta(k-f) \right) \alpha'  
\nonumber
\\
&+ \left(2 \alpha^2  - \alpha \left(4(k-f) - 2 \beta \right) + 4 \beta(k-f) \right) \alpha  \bigg] \, ,
\\ 
\mathcal{F}_{(10, 4)}^{(14)} =&+ \frac{f^2}{r^7} \bigg[ r^2 (\alpha + \beta) \left(8(k-f) - \alpha - \beta \right) \alpha'' - 4 r \left(\alpha^2 + \beta \left( 8 (k-f) - \beta \right) \right) \alpha' 
\nonumber
\\
&+ 8 (\beta-\alpha) \left(4(k-f) - \alpha -\beta \right) \alpha \bigg]
\\ 
\mathcal{F}_{(10, 4)}^{(15)} =&+ \frac{f^2 \alpha }{r^7} \bigg[ r^2(k-f) \alpha'' - (\alpha + \beta) \left(r \alpha' - \alpha + \beta \right) \bigg] \, ,
\\ 
\mathcal{F}_{(10, 4)}^{(16)} =&+ \frac{f^2  }{r^7} \bigg[r^2 \left(-\alpha^2 + \left(6(k-f) - \beta \right) \alpha + 2 \beta (k-f) \right) \alpha'' - 2 r \left(\alpha^2 + \alpha \beta + \beta(k-f)  \right) \alpha' 
\nonumber\\
&+ 4 \left(\alpha^2 - \left(2(k-f) - \beta \right) \alpha + 2 \beta (k - f) \right) \alpha \bigg] \, ,
\\
\mathcal{F}_{(10, 4)}^{(17)} =&+  \frac{f^2 \alpha }{r^7} \bigg[ r^2 \left( 4(k-f) - \alpha \right) \alpha'' - 4 \left(r \alpha' - \alpha + \beta \right) \beta \bigg]
\\ 
\mathcal{F}_{(10, 4)}^{(18)} =&+ \frac{f^2  }{r^7} \bigg[  r^2 \left( -4 r \left(3 \alpha + 3 \beta + 4 f - 12 k \right) \alpha' + 4 \left(3 \alpha^2 + 4 f \alpha - 12 k \alpha - 3 \beta^2 - 4 f \beta + 4 k \beta \right)  \right) \alpha'' 
\nonumber
\\
&+ 16 r^2 \left(  3 \alpha + 3 \alpha + 4 f - 9 k \right) \alpha'^2 - 2 r \left(47 \alpha^2 - \alpha \beta + 92 f \alpha - 28 k \alpha - 48 \beta^2 - 32 f \beta + 32 k \beta \right) \alpha' 
\nonumber
\\
&+ \left( 5 \alpha^2 - 178 \alpha \beta + 176 f \alpha + 464 k \alpha - 183 \beta^2 - 64 f \beta + 64 k \beta \right) \alpha \bigg) \bigg]
\\
\mathcal{F}_{(10, 4)}^{(19)} =&+ \frac{f^2  }{r^7} \bigg[96 f r^3 \alpha \alpha''' - 32 r^2 \left(-r f \alpha' + (19f - 9k) \alpha + \beta (k-f) \right) \alpha'' +  64 r^2 \left(3k-2f \right) \alpha'^2 
\nonumber
\\
&+  4 r \left(-163 \alpha^2 + \left(-148 k - 76 f - 163 \beta \right) \alpha + 32 \beta (k-f) \right) \alpha' 
\nonumber
\\
&- 2  \left( - 722 \alpha^2 - 2 \left( 35 \beta + 248 f + 392 k \right)  \alpha  + \beta \left(64(k-f) + 291 \beta \right) \right) \alpha\bigg] \, ,
\\
\mathcal{F}_{(10, 4)}^{(20)} =&+ \frac{f^2}{r^7} \bigg[ 6 r^3 f \left( r \alpha' - 2 \alpha \right) \alpha''' + 6 f r^4 \alpha''^2 - 4 \left( 4 r f \alpha'  + \left( 18 k - 13 f \right) \alpha + 5 \beta(k-f) \right)r^2 \alpha'' 
\nonumber
\\
&- 4 \left(3 k + 14 f \right) r^2 \alpha'^2 + 80 \left( \frac{\alpha^2}{16} + \left( \frac{23 k}{20} + \frac{41 f}{20} + \frac{\beta}{16} \right) \alpha  + \beta(k-f) \right) r \alpha' 
\nonumber
\\
&- 80 \left( \frac{7 \alpha^2}{32} + \left(-\frac{13 k}{10} +  \frac{23 f}{10} + \frac{5 \beta}{16} \right)\alpha + \beta \left( k -f  + \frac{3 \beta}{32} \right) \right) \alpha \bigg]
\\
\mathcal{F}_{(10, 4)}^{(21)} =&-\frac{3f^2}{r^7}\bigg[-\frac{1}{3} \left( -\alpha' f r -\frac{9 \alpha^ 2}{2}+\left(4k+2f-\frac{9\beta}{2}\right)\alpha + \beta (k-f)\right)r^2\alpha''\\
& -\frac{r^4 f \alpha \alpha'''' }{6} +r^3  \left( -\frac{\beta}{4}+k+\frac{f}{3}-\frac{\alpha}{4}\right) \alpha \alpha''' +2\left(k-\frac{2f}{3}\right)r^2\alpha''^2\\
&+\frac{4}{3}\left(-\frac{163\alpha^2}{32}-\left(\frac{77k}{8}+\frac{21 f}{8}+\frac{163\beta}{32} \right)\alpha+\beta(k-f) \right)r \alpha'\\
&-\frac{4\alpha}{3} \left(\frac{\beta' f r}{4}-\frac{361 \alpha^2}{64}-\left(\frac{69k}{4}+8f+\frac{35\beta}{32} \right)\alpha+\beta \left(k-\frac{3f}{4}+\frac{291\beta}{64} \right) \right)\bigg]\, .
\end{align}
As we can see, all densities but $\mathcal{F}_{(10, 0)}^{(1)}$ and $\mathcal{F}_{(10, 4)}^{(9)}$ involve linear combinations of terms proportional to either $\alpha(r)$, or $\beta(r)$, or their derivatives. Hence, for all those the Schwarzschild metric solves the corresponding equations of motion. The explicit form of the covariant densities is rather complicated in general, so we have preferred not to include the full list here. The corresponding expressions for  $\mathcal{F}_{(10, 0)}^{(1)}$ and $\mathcal{F}_{(10, 4)}^{(9)}$ read, respectively,
\begin{align}
\mathcal{L}^{(1)}_{10, 0} =&+\frac{1}{2160} \bigg[5 R^5 + 132 R  \left( R_{ab} R^{ab} \right)^2 + 18 R \left(R_{abcd} R^{abcd} \right)^2 - 272 R^2 \tensor{R}{_a^b_c^d}\tensor{R}{_b^e_d^f} \tensor{R}{_e^a_f^c} 
\nonumber
\\
&+ 10 R^2 \tensor{R}{_{ab}^{cd}} \tensor{R}{_{cd}^{ef}} \tensor{R}{_{ef}^{ab}} - 30 R^3 R_{ab}R^{ab} - 102 R R_{ab} R^{ab} R_{cdef} R^{cdef} 
\nonumber
\\
&+ 552 R_{ij} R^{ij} \tensor{R}{_a^b_c^d}\tensor{R}{_b^e_d^f} \tensor{R}{_e^a_f^c} - 156 R_{ijkl} R^{ijkl} \tensor{R}{_a^b_c^d}\tensor{R}{_b^e_d^f} \tensor{R}{_e^a_f^c}\bigg]
\\
\mathcal{L}^{(9)}_{10, 4} =& -\frac{1113943}{20864} \tensor{C}{_{abcd}} \tensor{C}{^{abcd}}  \tensor{C}{^{efgh;i}} \tensor{C}{_{efgh;i}}
+ \frac{19309071}{39446} R_b^a R_c ^b \tensor{R}{_{ae}^{cd}} \tensor{R}{_{gh}^{ef}} \tensor{R}{_{df}^{gh}}  
\nonumber
\\
&+ \frac{2168502179}{4733520} R_c^a R_d ^b \tensor{R}{_{ef}^{cd}}\tensor{R}{_{gh}^{ef}} \tensor{R}{_{ab}^{gh}} - \frac{23092199}{10758} R_b^a \tensor{R}{_{ad}^{bc}} \tensor{R}{_{fh}^{de}}\tensor{R}{_{c i}^{fg}} \tensor{R}{_{eg}^{hi}}  
\nonumber
\\
&+ \frac{7605694303}{4733520} R_b^a \tensor{R}{_{de}^{bc}}\tensor{R}{_{cf}^{de}} \tensor{R}{_{hi}^{fg}} \tensor{R}{_{ag}^{hi}}  + \frac{2051116779}{788920} \tensor{R}{_{cd}^{ab}} \tensor{R}{_{eg}^{cd}} \tensor{R}{_{ai}^{ef}} \tensor{R}{_{fj}^{gh}} \tensor{R}{_{bh}^{ij}} 
\nonumber
\\
&- \frac{6886022969}{2366760}  \tensor{R}{_{ce}^{ab}} \tensor{R}{_{af}^{cd}} \tensor{R}{_{gi}^{ef}} \tensor{R}{_{bj}^{gh}} \tensor{R}{_{dh}^{ij}}  + \frac{176696887}{215160} \tensor{R}{_{ce}^{ab}} \tensor{R}{_{fg}^{cd}} \tensor{R}{_{hi}^{ef}} \tensor{R}{_{aj}^{gh}} \tensor{R}{_{bd}^{ij}} 
\nonumber
\\
&+ \frac{237411}{7172} R_{ab}R^{ab} \tensor{R}{^{pqrs;t}} \tensor{R}{_{pqrs;t}} - \frac{388530029}{11360448} R_{abcd} R^{abcd} R^{;p}R_{;p}  + \frac{649013}{7824} R_{abcd} R^{abcd} R^{pq;r}R_{pq;r} 
\nonumber
\\
&+ \frac{472435}{43032} R_{abcd} R^{abcd} R^{pq;r} R_{pr;q} - \frac{1802455}{57376} R_{abcd} R^{abcd} R^{pq;rs} R_{p r q s} 
\nonumber
\\
&- \frac{1065937721}{2840112} R_{pqrs} R^{pqrs} \tensor{R}{_a^c_b^d} \tensor{R}{_c^e_d^f}\tensor{R}{_e^a_f^b} + \bigg[ \frac{1535482063}{22720896}R^{;a} R_{;a}  - \frac{34589843}{631136}  R^{pq;r} R_{pq;r}  
\nonumber
\\
& - \frac{90238183}{946704} R^{pq;r} R_{pr;q}    + \frac{21168615}{2524544} R^{pqrs;t} R_{pqrs;t}   + \frac{315857975}{1893408} R^{;pq} R_{pq}   - \frac{6185788187}{17040672} R_a^b R_b^c R_c^a 
\nonumber
\\
&- \frac{939340}{177507} \tensor{R}{_a^c_b^d} \tensor{R}{_c^e_d^f}\tensor{R}{_e^a_f^b}  - \frac{26416471}{516384} \tensor{R}{_{ab}^{cd}} \tensor{R}{_{cd}^{ef}} \tensor{R}{_{ef}^{ab}}\bigg] \Box R 
\nonumber
\\
&+ \bigg[ \frac{1612029697}{1893408}  R^{pq}R_p^r \Box R_{qr} + \frac{74535679}{118338} R^{p q} R^{r s} R_{pq ; rs} + \frac{48934355}{236676} R^{p q ; rs} \tensor{R}{^t_p^u_r} R_{t q u s}
\nonumber
\\
&+\frac{23734313}{59169} R^{pq} R^{rs;qt} R_{p r s t}+\frac{35456237}{946704} R^{pq} R^{r s t u} R_{r s t u ;p q}-\frac{44597992 }{59169} R^{pq} R^{rs} R_{p r ; q s} 
\nonumber
\\
&-\frac{619200179 }{1420056} R^{;p q} R^{r s} R_{p r q s}-\frac{315857975 }{1893408} R^{p q} \nabla_q \nabla_p \Box R +\frac{90238183 }{473352} R^{p q ; r} \nabla_q \Box R_{p r}
\nonumber
\\
&-\frac{21168615}{315568} R^{p q; rst}R_{p r q s ; t}+\frac{293954069 }{2840112}R^{;pq} \Box R_{p q} +\frac{1588811801 }{5680224} R R^{p q ; r s} R_{p r q s} 
\nonumber
\\
&+\frac{15025369}{19723} R^{pq} \Box R^{r s} R_{p r q s}+\frac{1535482063 }{11360448} R^{;pq} R_{;pq} +\frac{90238183 }{473352} R^{p q ; r s} R_{p r ; q s}
\nonumber
\\
&+\frac{26525693 }{258192} R^{;p q} R_p^r R_{q r}+\frac{34589843 }{315568} R^{p q; r s} R_{p q; rs}-\frac{681365365 }{946704}R R^{pq} \Box R_{pq}
\nonumber
\\
&+\frac{34589843 }{315568} R^{p q; r} \nabla_r \Box R_{p q}+\frac{293954069 }{1420056} R^{;p q r} R_{pq;r} +\frac{98790361 }{473352} R^{pq;r} \tensor{R}{_p^{s;t}} R_{q r s t}
\nonumber
\\
&-\frac{333105233 }{315568} R^{p q; r} \tensor{R}{_p^{s;t}} R_{q s r t}+\frac{17765777 }{86064} R^{pq;r} \tensor{R}{^{st}_p^u} R_{s t q r ;u}
-\frac{425439281 }{946704} R^{p q ;r} \tensor{R}{^s_p^{tu}} R_{s q t r ;u} 
\nonumber
\\
&+\frac{16960493 }{187776} R^2 \Box R +\frac{421946281}{315568} R^{p q r s} \tensor{R}{^{tuv}_{p;q}} R_{tuvr;s} -\frac{21168615 }{1262272} R^{p q r s ; t u} R_{p q r s ; t u} 
\nonumber
\\
&+\frac{238362363 }{631136} R^{p q r s} \tensor{R}{^{tuv}_{p;r}} R_{tuvq;s}-\frac{9210385 }{315568} R^{p q r s} \tensor{R}{^t_p^u_r^{;v}} R_{t q u s ;v} 
\nonumber
\\
&+\frac{298907053 }{1893408} \tensor{R}{^{m n r s}} \tensor{R}{_m^d_r^g} \tensor{R}{_d^c_g^i} R_{n c s i}-\frac{298907053 \mathit{L85}}{7573632} \tensor{R}{^{m n r s}} \tensor{R}{_{m n} ^{ d g}} \tensor{R}{_{d g}^{c i}} \tensor{R}{_{r s c i}}\bigg] R\, .\label{L1049}
\end{align}

\subsection{Black hole solutions}
In this section we present the first examples of black hole solutions to GQT theories with covariant derivatives. These are continuous deformations of the Schwarzschild metric and solve the equations of motion of Einstein gravity coupled to the two non-trivial ten-derivative GQT densities presented above.  Note that in appendix \ref{8dpure} we construct additional  (analytic) examples of non-Schwarzschild solutions in the case of eight-derivative GQT densities. However, those correspond to the less interesting case in which we consider a linear combination of GQT densities but no Einstein gravity term. 

Let us then consider the gravitational Lagrangian given by
\begin{equation}
\mathcal{L}=\frac{1}{16\pi G} \left[R +  \frac{5 \mu}{4} \mathcal{L}_{(10, 0)}^{(1)}  + \frac{\lambda}{8} \mathcal{L}_{(10, 4)}^{(9)}  \right]\, ,
\end{equation}
where the explicit form of the ten-derivative densities can be found in Section~\ref{sec:4dTheories} and for convenience we redefined the gravitational couplings in terms of two new parameters, $c_{(1)}\equiv 5\mu/4$ and $c_{(9)}\equiv \lambda/8$.
For this theory, the  field equations for the SSS ansatz reduce to\footnote{Henceforth we set $G = 1$ in this section.}
\begin{align}
\mathcal{F}_{(2, 0)}^{(1)} +  \frac{5 \mu}{4} \mathcal{F}_{(10, 0)}^{(9)}  + \frac{\lambda}{8} \mathcal{F}_{(10, 4)}^{(9)} = 4 G M \, .
\end{align}
where again the individual contributions can be found in Section~\ref{sec:4dTheories}. 

We present a relatively brief analysis of the solution. Working perturbatively at large $r$ and to linear order in the coupling constants, we find the correction to the Schwarzschild solution to be
\be 
f(r) = 1 - \frac{2M}{r} - \frac{4752 \lambda M^3 }{r^{11}}  + \frac{ 90 (\mu + 208 \lambda) M^4}{r^{12}} - \frac{2 (83 \mu + 9236 \lambda ) M^5 }{r^{13}} + \cdots \, .
\ee
This makes clear already that the two densities correct the solution in inequivalent ways. On the other hand, since we are interested in black hole solutions, we consider a near-horizon solution of the field equations. In this regime, we write an expansion for the metric function
\be 
f(r) = 4 \pi T (r-r_h) + \sum_{i=2}^\infty a_i (r-r_h)^i \, ,
\ee
and expand the field equations as $r \to r_h$. Remarkably, the usual characteristic property of GQT theories continues to hold. The first two terms in the near horizon expansion of the field equations suffice to fully determine the black hole thermodynamics analytically. These equations read,
\begin{align}
M &= \frac{r_h}{2}  + \frac{8 \pi^4 T^4 \mu \left(5 + 8 \pi  r_h T  \right)}{r_h^3} - \frac{2 \pi^2 T^2 \lambda \left[1 + 12 \pi r_h T + 16 \pi^2 r_h^2 T^2 \right]}{r_h^5} \, ,
\\
0 &= 1 - 4 \pi r_h T + \frac{16 \pi^4 T^4 \mu \left(5 + 4 \pi  r_h T  \right)}{r_h^4} + \frac{4 \pi^2 T^2 \lambda  \left(1 + 4 \pi r_h T \right) \left(5 + 4 \pi r_h T \right)}{r_h^6} \, .
\label{thermo_eqns}
\end{align}
The first equation above expresses the mass $M$ as a function of the temperature $T$ and the horizon radius $r_h$, while the second determines the temperature as a function of the horizon radius.  At the next order in the near horizon expansion, the parameters $a_2$ and $a_3$ appear, the latter linearly. The higher-order terms in the expansion can be solved for $a_n$ $(n \ge 3)$ in terms of a single free parameter $a_2$. This is exactly the same behaviour typically seen for GQT theories with second-order integrated equations \cite{PabloPablo2,Hennigar:2017ego,PabloPablo4}. Here, one of the theories has second-order integrated equations, while the other has third-order. Nonetheless, we find that this does not change the usual picture for the near-horizon solution.

We wish to understand the effects of the corrections to the thermodynamics of the Schwarzschild black hole. The  near-horizon equations give us the mass and temperature, and so only the entropy remains. Computing the Wald entropy \cite{Wald:1993nt,Iyer:1994ys} for this theory is rather involved, so we instead use the first law itself to determine the entropy.  Regarding the temperature as a function of horizon radius $T = T(r_h)$, we can obtain an expression for $\diff M$ in terms of the temperature, its first derivative, and $r_h$. The first law tells us that $\diff M/T$ must be an exact differential. By adding $-1/(2 T )$ times the constraint~\eqref{thermo_eqns} to the expression for $\diff M/T$ we can confirm that it is exact, and therefore can be directly integrated. This gives for the entropy
\be 
S = \pi r_h^2 \left[1 + \frac{80 \pi ^3 T^3 \mu \left(2 + 3 \pi r_h T  \right)}{3 r_h^5} - \frac{4 \pi T \lambda \left(3 + 27 \pi r_h T + 32 \pi^2 r_h^2 T^2 \right)}{r_h^7}  \right] \, .
\ee
By construction, the thermodynamic quantities satisfy the first law $\diff M = T\diff S$. It should be possible to verify this relation by a direct computation of the Wald entropy, although this would be rather challenging computationally. 

\begin{figure}[t]
\centering
\includegraphics[width=0.65\textwidth]{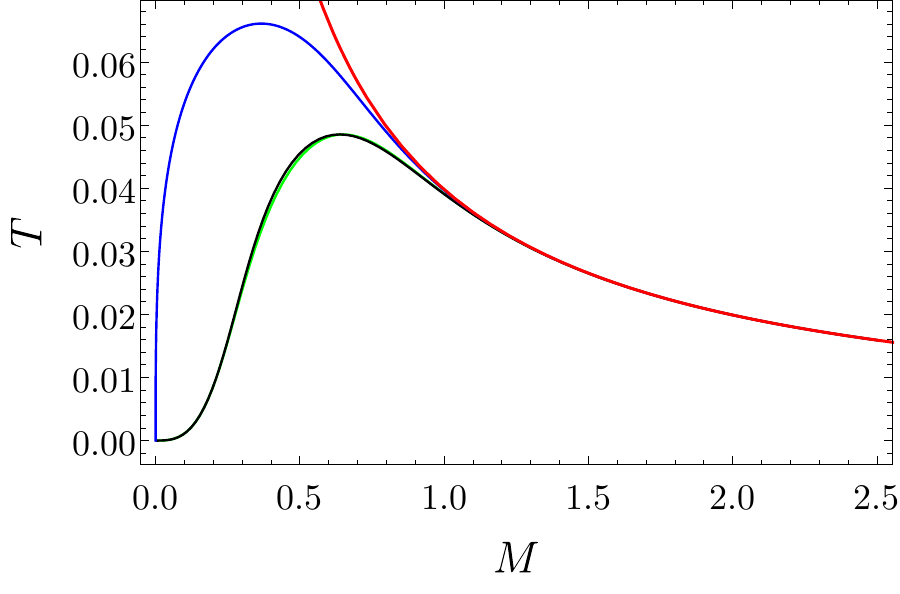}
\caption{We show the effect of the higher derivative corrections on the temperature versus mass relation, relative to the Einstein gravity case (red curve). The coloured curves show the temperature versus mass for $\lambda = 1$ and $\mu = -1$ (black curve), $\lambda = 1$ and $\mu = 0$ (green curve) and $\lambda = 0$ and $\mu = -1$ (blue curve). The corrections controlled by $\lambda$ generically dominate in the small black hole regime. Dimensionful parameters are measured in units of the coupling constants.}
\label{temp-v-rh}
\end{figure}

It is interesting to compare the effects of the $\mu$-controlled corrections (which correspond to the previously known family of GQT theories first studied in~\cite{PabloPablo4}) with the new $\lambda$-controlled higher-derivative corrections. For $\lambda > 0$ we plot the relationship between temperature and mass for these black holes in Figure~\ref{temp-v-rh}. The plot compares the cases with $(\mu, \lambda) \in \left\{(-1,1), (0, 1), (-1, 0) \right\}$.\footnote{In this case, that $\mu < 0$ follows from the general analysis of~\cite{PabloPablo4}.} In all cases there is a maximum value of the temperature of the corrected black holes. Below this temperature the specific heat becomes positive.

Let us explore the features of the small black holes, keeping in mind that for sufficiently small black holes additional corrections would be expected to become important.\footnote{For sufficiently small horizon radius, the entropy becomes negative. However, the entropy can be shifted by an arbitrary constant by adding a topological Gauss-Bonnet term to the action. So the region of negative entropy is not worrisome.} Generically the $\lambda$-controlled theory dominates in the small black hole regime. The two theories give rise to different scaling behaviour for the temperature of small black holes. The $\mu$-controlled theory has $T \sim M^{1/3}$ as $r_h \to 0$, while the $\lambda$-controlled theory has $T \sim M^3$.  Interestingly, the infinite class of GQT theories based on polynomial curvature invariants as studied in~\cite{PabloPablo4, Bueno:2019ycr} uniformly display a temperature scaling of $T \sim M^{1/3}$ for small black holes. Similarly, in the case of exclusively polynomial invariants, the modified Smarr relation $M=\frac{2}{3} TS$ universally holds for small black holes for general GQT theories \cite{PabloPablo4}. When only $\lambda$ is active, we find yet another version of the Smarr relation in this regime, namely, $M=TS$. Hence, the $\lambda$-controlled theory, deviating from these patterns, is a unique and noteworthy instance. It is tempting to speculate with the possibility that the universal patterns identified in the case of polynomial theories may have universal counterparts for theories involving covariant derivatives. Additionally, these deviations from the purely polynomial case suggest that terms with covariant derivatives might play a pivotal role in understanding characteristics of small black holes, such as their evaporation.

\begin{figure}[t]
\centering
\includegraphics[width=0.65\textwidth]{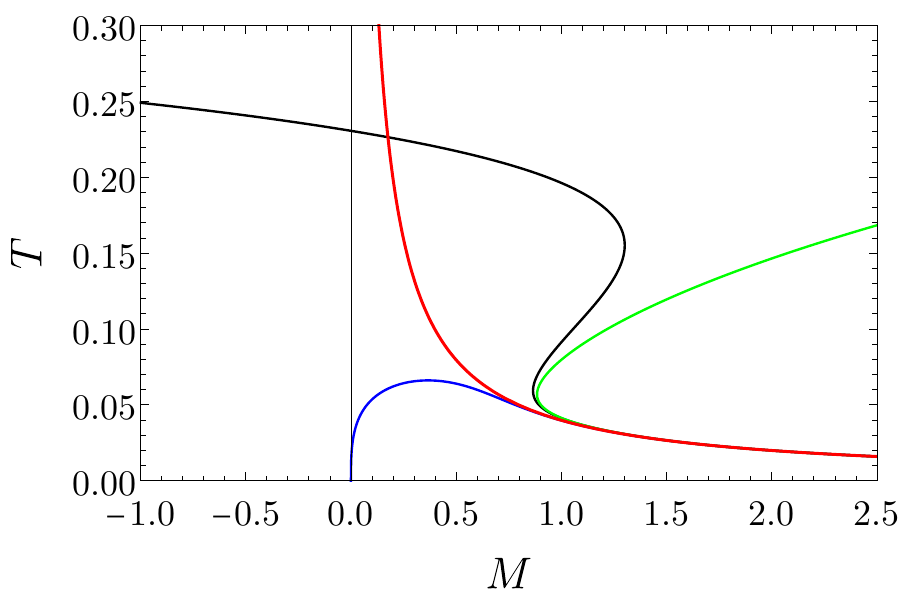}
\caption{We show the effect of the higher derivative corrections on the temperature versus mass relation, relative to the Einstein gravity case (red curve).  The coloured curves show the temperature versus mass for $\lambda = -2$ and $\mu = -1$ (black curve), $\lambda = -2$ and $\mu = 0$ (green curve) and $\lambda = 0$ and $\mu = -1$ (blue curve). When both higher-derivative couplings are active, the mass is unbounded from below with $M\to - \infty$ as $r_h \to 0$. Dimensionful parameters are measured in units of the coupling constants.}
\label{temp-v-rh2}
\end{figure}

We consider next the situation with $\lambda < 0$, which displays some significant differences relative to what we have just seen. In the previous case, the qualitative behaviour of the two theories was similar, here they are different --- see Figure~\ref{temp-v-rh2}. The $\lambda$-controlled theory, for negative coupling $\lambda$, has a minimum black hole size and mass when $\mu$ is strictly zero. However, for any finite value of $\mu$ the situation is completely different and qualitatively similar to the black curve shown in the figure. First, let us note that when both $\lambda$ and $\mu$ and  are negative we have
\be 
T = \frac{x + \sqrt{x(x+1)}}{2 \pi} \frac{1}{r_h} + \cdots  \quad \text{with} \quad x \equiv  \frac{\lambda}{\mu} \quad \text{as} \quad r_h \to 0. 
\ee
So small black holes have large temperature (for $x = 1/8$ the relationship is the same as in Einstein gravity). However, for sufficiently small $r_h$ the black hole mass ultimately becomes negative and approaches $M \to -\infty$ as $r_h \to 0$. As such, this branch of solutions exhibits rather pathological behaviour, as the small black holes exhibit large negative masses. This results in an order of limits issue, and flat space is not recovered as $r_h \to 0$. 

We postpone a more detailed and systematic study of the thermodynamic properties of the black holes of GQTs involving covariant derivatives for future work.


\section{Brane-world gravities}\label{bW}
In this section we consider a different class of  gravitational theories constructed from contractions of the Riemann tensor and its covariant derivatives, namely, brane-world gravities \cite{Randall:1999vf}. We find a closed expression for the quadratic-order action, which involves a combination of inverse polynomials and Bessel functions of the Laplace operator. Using this, we analyze the linearized spectrum of brane-world gravities. 
We generically find infinite towers of massive ghost-like gravitons. In five dimensions we find an additional tachyonic mode, whereas in seven dimensions we find two extra modes with complex squared masses which are conjugate of each other.  On the other hand, both in four and six dimensions, we find infinite towers of pairs of modes with conjugate complex squared masses.    

In the context of $(D+1)$-dimensional Einstein-AdS gravity, the insertion of a co-dimension one brane near the AdS boundary gives rise to an effective theory for the brane induced metric coupled to a cut-off CFT. The gravitational theory involves an infinite series of higher-derivative terms built from the Riemann tensor and its covariant derivatives. Let us quickly review how this comes along. Consider the action of Einstein gravity coupled to a cosmological constant in general dimensions
\begin{equation}
I=\frac{1}{16\pi G} \left[\int_{\mathcal{M}} \diff^ {D+1}X \sqrt{-G} \left(R[G]+\frac{D(D-1)}{\ell^2} \right)+2\int_{\partial \mathcal{M}}\diff^D x \sqrt{-g} \, K \right]\, .
\end{equation}
Inserting a brane near the AdS$_{(D+1)}$ boundary amounts to introducing an additional term of the form
\begin{equation}
I_b= - T \int_{\rho= \varepsilon} \diff^D x \sqrt{-g}\, ,
\end{equation}
where $\varepsilon \ll \ell$, $T$ is the brane tension and $\rho$ is the holographic coordinate which we can use to write the bulk metric in a Fefferman-Graham expasion \cite{2007arXiv0710.0919F}
\begin{equation}
G_{\mu\nu}\diff x^{\mu}\diff x^{\nu}=\frac{\ell^ 2}{4\rho^ 2} \diff \rho^2 + \frac{\ell^ 2}{\rho} \hat g_{ij} (\rho,x)\diff x^ i\diff x^ j\, .
\end{equation}
The total action, $I+I_b$, can be alternatively written as a sum of a gravitational action for the brane induced metric and an quantum effective action of a CFT living on the brane, namely,
\begin{equation}
I+I_b = I_{\rm bgrav}+ I_{\rm CFT}\, .
\end{equation}
The defining property of the induced theory of gravity on the brane $I_{\rm bgrav}$ ---which follows from the Israel junction conditions in the AdS bulk \cite{Israel:1966rt}--- is that its equations of motion,
\begin{equation}
\Pi^{ab}\equiv \frac{2}{\sqrt{-g}}\frac{\delta}{\delta g_{ab}}\int \diff ^{D}x\sqrt{-g}\mathcal{L}\, ,
\end{equation}
satisfy the identity \cite{Kraus:1999di}
\begin{equation}\label{eq:Piid1}
\frac{1}{D-1}\Pi^2-\Pi_{ab}\Pi^{ab}=\frac{D(D-1)}{\ell^2}+R
\end{equation}
in the case of a tensionless brane.
The claim is that there exists a unique theory with this property. This can be reformulated as the fact that there is a unique conserved tensor (\textit{i.e.}, satisfying the identity $\nabla_{a}\Pi^{ab}=0$) built out of the intrinsic metric $g_{ab}$ that satisfies this relation.  Both facts, the existence and uniqueness of this theory, are remarkable. 

This theory has a definite value of the cosmological constant, but it is possible to shift this value by adding a tension to the brane. Introducing a non-vanishing tension amounts to performing
\begin{equation}
\Pi_{ab}\rightarrow \Pi_{ab}+T g_{ab}\, .
\end{equation}
in \eqref{eq:Piid1}.
We fix the brane tension so that the theory has a vanishing cosmological constant, as we will be interested in asymptotically flat solutions. This is achieved for $T=(D-1)/\ell$, so that the equation satisfied by the new $\Pi_{ab}$ reads
\begin{equation}
\Pi=\frac{\ell}{2}\left[R+\Pi_{ab}\Pi^{ab}-\frac{1}{D-1}\Pi^2\right]\, .
\end{equation}
In order to solve this equation, we assume that the Lagrangian allows for a derivative expansion of the form
\begin{equation}
\mathcal{L}=\sum_{n=1}^{\infty} \ell^{2n-1}\mathcal{L}_{(n)}\, ,
\end{equation}
and similarly 
\begin{equation}
\Pi^{ab}=\sum_{n=1}^{\infty} \ell^{2n-1}\Pi^{ab}_{(n)}\, .
\end{equation}
Then, we get \cite{Bueno:2022log}
\begin{align}
\Pi_{(1)}&=\frac{R}{2}\, ,\\
\Pi_{(n)}&=\frac{1}{2}\sum_{i=1}^{n-1}\left[\Pi_{(i)\, ab}\Pi^{ab}_{(n-i)}-\frac{1}{D-1}\Pi_{(i)}\Pi_{(n-i)}\right]\, , \quad n\ge 2\, .
\label{Pinrecursive1}
\end{align}
The other ingredient we need to solve this recursive relation is \cite{Kraus:1999di}
\begin{equation}\label{PiL}
\Pi_{(n)}=(D-2n)\mathcal{L}_{(n)}+\text{total derivative}\, .
\end{equation}
Since the total derivatives are irrelevant for the Lagrangian, this allows us to get $\mathcal{L}_{(n)}$ from the trace of the equation of motion $\Pi_{(n)}$.
Thus, we get
\begin{equation}
\mathcal{L}_{(1)}=\frac{R}{2(D-2)}\, ,\quad \Pi_{(1)\, ab}=-\frac{1}{D-2}G_{ab}\, .
\end{equation}
In a similar fashion, this process allows us to generate all the Lagrangian densities $\mathcal{L}_{(n)}$. Observe that all of these Lagrangians will be of the form
\begin{equation}
\mathcal{L}=\mathcal{L}(R_{ab},\nabla_{c}R_{ab}, \nabla_{c}\nabla_{d}R_{ab}, \ldots)\, ,
\end{equation}
since Riemann curvature appears nowhere in the process. The quadratic and cubic densities read, respectively, \cite{Kraus:1999di,Emparan:1999pm,Balasubramanian:1999re,Papadimitriou:2004ap,Papadimitriou:2010as,Elvang:2016tzz,Bueno:2022log,Anastasiou:2020zwc}
\begin{align}
    \mathcal{L}_{(2)}=   &  +\frac{1}{2(D-2)^2(D-4)} \left[ R_{ab}R^{ab} - \frac{D}{4(D-1)} R^2 \right] ,  \label{quada} \\ \label{cubi}
    \mathcal{L}_{(3)} 
   =& - \frac{1}{(D-2)^3(D-4)(D-6)} \Bigg[ \frac{3D+2}{4(D-1)}RR_{ab}R^{ab} - \frac{D(D+2)}{16(d-1)^2}R^3 - 2R^{a}_b R^{b}_c R^{c}_a \nonumber \\
    &\qquad\quad + \frac{D}{2(D-1)}R^{ab}\nabla_{a}\nabla_{b} R + 2 R^{ab}\nabla^c\nabla_b R_{ac} - R^{ab} \Box R_{ab} + \frac{1}{2(D-1)} R \Box R \Bigg].
\end{align}
where already at cubic order we start seeing the appearance of covariant derivatives of the Ricci tensor. Explicit formulas for the quartic and quintic terms appear in \cite{Bueno:2022log}.

\subsection{Quadratic action}
We are interested in studying the linearized equations of these theories around the Minkowski vacuum. As we have seen, the only higher-derivative terms that contribute to the linearized equations are those quadratic in the curvature (but with an arbitrary number of covariant derivatives) and, therefore, the only possible quadratic Lagrangians are $R\Box^{n} R$ and $R^{ab}\Box^n R_{ab}$.  Thus, at order $2n$ in derivatives, we will necessarily have
\begin{equation}\label{Ln}
\mathcal{L}_{(n)}=\alpha_{n} R\Box^{n-2} R+\beta_{n}R^{ab}\Box^{n-2}R_{ab}+\mathcal{O}(R^3)\, .
\end{equation} 
Our goal is to determine the coefficients $\alpha_{n}$ and $\beta_{n}$, for which we will use \req{Pinrecursive1}. First of all, in order to evaluate the left-hand-side of \req{Pinrecursive1}, we use \req{PiL}, so that we get
\begin{equation}\label{PiL2}
\Pi_{(n)}=(D-2n)\left(\alpha_{n} R\Box^{n-2} R+\beta_{n}R^{ab}\Box^{n-2}R_{ab}\right)+\ldots\, .
\end{equation}
Now we must evaluate the right-hand-side. The case $n=2$ must be considered independently, and it yields
\begin{equation}
\Pi_{(2)}=\frac{1}{2}\left[\Pi_{(1)\, ab}\Pi^{ab}_{(1)}-\frac{1}{D-1}\Pi_{(1)}^2\right]=-\frac{D}{8(D-1)(D-2)^2}R^2+\frac{1}{2(D-2)^2}R^{ab}R_{ab}\, ,
\end{equation}
so that we identify
\begin{equation}
\alpha_{2}=-\frac{D}{8(D-1)(D-2)^2(D-4)}\, ,\quad \beta_{2}=\frac{1}{2(D-2)^2(D-4)}\, .
\end{equation}
Now, for $n\ge 3$ we have
\begin{align}\notag
\Pi_{(n)}&=\Pi_{(1)\, ab}\Pi^{ab}_{(n-1)}-\frac{1}{D-1}\Pi_{(1)}\Pi_{(n-1)}+\frac{1}{2}\sum_{i=2}^{n-2}\left[\Pi_{(i)\, ab}\Pi^{ab}_{(n-i)}-\frac{1}{D-1}\Pi_{(i)}\Pi_{(n-i)}\right]\\
&=-\frac{R_{ab}\Pi^{ab}_{(n-1)}}{D-2}+\frac{R\Pi_{(n-1)}}{2(D-1)(D-2)}+\frac{1}{2}\sum_{i=2}^{n-2}\left[\Pi_{(i)\, ab}\Pi^{ab}_{(n-i)}-\frac{1}{D-1}\Pi_{(i)}\Pi_{(n-i)}\right] \, .\label{Pin2}
\end{align}
In order to evaluate this expression we need the equations of motion $\Pi_{(n)\, ,ab}$. Notice that we will compare the resulting expression with \req{PiL2}, which is quadratic in the curvature. Now, \req{Pin2} is already quadratic in the equations of motion, and this means that, in order to obtain the terms that are quadratic in the curvature we only need to obtain the terms in the equations of motion that are linear in the curvature. Fortunately, all of these come from the term
\begin{equation}
-4\nabla^{c}\nabla^{e}P_{acbe}\subset \Pi_{(n)\, ,ab}\, ,\quad \text{where}\quad P_{acbe}=\frac{\delta \mathcal{L}}{\delta R^{abcd}}\, .
\end{equation}
For a theory that only depends on Ricci curvatures this can be expressed as 
\begin{equation}
\Pi_{(n)\, ,ab}=-2 g_{ab}\nabla^{c}\nabla^{e} P_{ce}-2\Box P_{ab}+4\nabla^{c}\nabla_{(a}P_{b)c}+\ldots\, ,\quad  \text{where}\quad P_{ab}=\frac{\delta \mathcal{L}}{\delta R^{ab}}\, .
\end{equation}
Thus, for the Lagrangians (\ref{Ln}) we get
\begin{align}
\Pi_{(n)\, ,ab}&=-(4\alpha_n+\beta_n)g_{ab}\Box^{n-1}R+2(2\alpha_n+\beta_n)\nabla_{a}\nabla_{b}\Box^{n-2}R-2\beta_{n}\Box^{n-1}R_{ab}+\ldots\, ,\\
\Pi_{(n)}&=-\left(4(D-1)\alpha_n+D \beta_n\right)\Box^{n-1}R+\ldots \, 
\end{align}
Then, we can use these expressions to evaluate \req{Pin2}, and after some simplifications we find
\begin{equation}
\Pi_{(n)}=2\left(-\frac{D}{4(D-1)}R\Box^{n-2} R+R^{ab}\Box^{n-2}R_{ab}\right)\left[\frac{\beta_{n-1}}{(D-2)}+\sum_{i=2}^{n-2}\beta_{i}\beta_{n-i}\right]+\ldots\, ,
\end{equation}
where the ellipsis also contain total derivatives that arise when rearranging the derivatives. Therefore, comparing with \req{PiL2}, we conclude that 
\begin{equation}
\alpha_{n}=-\frac{D}{4(D-1)}\beta_{n}\, ,
\end{equation}
while $\beta_{n}$ satisfies the recursive relation 
\begin{equation}
\beta_{n}=\frac{2}{(D-2n)}\left[\frac{\beta_{n-1}}{(D-2)}+\sum_{i=2}^{n-2}\beta_{i}\beta_{n-i}\right]\, .
\end{equation}
We can transform this recursive relation into a differential equation by introducing the generating function 
\begin{equation}\label{smallf}
f(x)=\sum_{n=2}^{\infty}\beta_{n}x^{2n-D}\, .
\end{equation}
By taking the derivative and using the recursive relation for $\beta_{n\ge 3}$, we have 
\begin{equation}
\begin{aligned}
f'(x)&=\sum_{n=2}^{\infty}(2n-D)\beta_{n}x^{2n-D-1}=(4-D)\beta_{2} x^{3-D}+2\sum_{n=3}^{\infty}\left[\frac{\beta_{n-1}}{D-2}+\sum_{i=2}^{n-2}\beta_{i}\beta_{n-i}\right]x^{2n-D-1}\\
&=(4-D)\beta_{2} x^{3-D}-\frac{2}{D-2}x f(x)-2 x^{D-1} f(x)^2\, .
\end{aligned}
\end{equation}
Now, the action can in fact we written in terms of this function. The full action (at quadratic order) reads
\begin{align}\notag
I_{\rm bgrav}^{(2)}&=\frac{1}{16\pi G_{D+1}}\int  \diff^{D}x\sqrt{-g}\left[\frac{\ell}{2(D-2)}R+\sum_{n=2}^{\infty}\beta_{n}\ell^{2n-1}\left(R^{ab}\Box^{n-2}R_{ab}-\frac{D}{4(D-1)}R\Box^{n-2} R\right)\right]\, ,\\
&=\frac{1}{16\pi G_{D}}\int  \diff^{D}x\sqrt{-g}\left[R+\ell^2 R^{ab}F\left(\ell^2\Box\right)R_{ab}-\frac{D}{4(D-1)}\ell^2 RF\left(\ell^2\Box\right)R\right]\, ,\label{totalaction}
\end{align}
where 
\begin{equation}\label{bigF}
F(\ell^2\Box)=2(D-2)\sum_{n=2}^{\infty}\beta_{n}\left(\ell^2\Box\right)^{n-2}\, ,
\end{equation}
and $G_{D}=2(D-2)G_{D+1}/\ell$. We see that this $F$ is related to $f$ in \req{smallf} by 
\begin{equation}
f(x)=\frac{1}{2(D-2)}x^{4-D}F(x^2)\, .
\end{equation}
Thus, $F(x)$ satisfies the equation
\begin{equation}
F'(x)=(D-4)\frac{F(x)-F(0)}{2x}-\frac{1}{2(D-2)}\left(2F(x)+x F(x)^2\right)\, ,
\end{equation}
where 
\begin{equation}
F(0)=2(D-2)\beta_2=\frac{1}{(D-2)(D-4)}\, .
\end{equation}
Remarkably, this differential equation allows for a general solution in terms of Bessel functions. We find that the appropriate solution, that corresponds to the summation of the series \req{bigF}, is given by 
\begin{equation}
F_{D}(x)=\frac{D(D-2)}{x^2}-\frac{1}{x}-\frac{(D-2)Y_{\frac{D+2}{2}}\left(\sqrt{x}\right)}{x^{3/2}Y_{\frac{D}{2}}\left(\sqrt{x}\right)}\, ,
\end{equation}
where $Y_{k}$ are the Bessel functions of the second kind. Inserting $F_{D}(\ell^2 \Box)$ in \req{totalaction} we obtain our final expression for the quadratic action of the brane-world theory in general dimensions.

Despite the singular appearance of this function at $x=0$, it is actually analytic around that point for odd $D$. In fact, for odd $D$, $F_{D}$ can actually be written in terms of trigonometric functions. We have
\begin{align}
F_{3}(x)&=-\frac{\sin \left(\sqrt{x}\right)}{x \sin \left(\sqrt{x}\right)+\sqrt{x} \cos \left(\sqrt{x}\right)}\approx -1 +\frac{2x}{3}-\frac{7x^2}{15}+\frac{34x^3}{105}+\dots \, ,\\
F_{5}(x)&=\frac{\cos \left(\sqrt{x}\right)}{(3-x) \cos \left(\sqrt{x}\right)+3 \sqrt{x} \sin \left(\sqrt{x}\right)}\approx \frac{1}{3}-\frac{2x}{9}+\frac{x^2}{27}+\frac{2x^3}{405}+\dots\, ,\\
F_{7}(x)&=-\frac{\sqrt{x} \sin \left(\sqrt{x}\right)+\cos \left(\sqrt{x}\right)}{(x-15) \sqrt{x} \sin \left(\sqrt{x}\right)+3 (2 x-5) \cos \left(\sqrt{x}\right)} \approx \frac{1}{15}+\frac{2x}{75}-\frac{13x^2}{1125}+\frac{22x^3}{16875}+\dots\, ,
\end{align}
where we included the first terms in the expansions around $x=0$.
On the other hand, in even $D\geq 4$, the expansion around $x=0$ contains logarithmic divergences, which are the counterpart of the $1/(D-2n)$ divergences in the definition of these theories. For instance, for $D=4$ one finds
\begin{equation}
F_4(x)=\frac{8}{x^2}-\frac{1}{x}-\frac{2 Y_{3}(\sqrt{x})}{x^{3/2} Y_2(\sqrt{x})}\approx \frac{1}{4}\left[-2\gamma_{\rm E}-\log(x/4) \right]+\frac{1}{8}\left[ -1+\gamma_{\rm E}+\log(x/4)\right]x + \dots
\end{equation}
where $\gamma_{\rm E}$ is the Euler-Mascheroni constant. Finally, the $D=2$ case is a bit different, as it simply yields 
\begin{equation}
F_2(x)=-\frac{1}{x}\, ,
\end{equation}
which means that the corresponding quadratic action is proportional to the Polyakov induced-gravity action \cite{Polyakov:1987zb} ---see also \cite{Chen:2020uac}.

\subsection{Linearized equations and modes}
It is obvious from \req{totalaction} that the brane-world theory belongs to the class of theories which satisfy condition (\ref{keyko}), as in this case we have $F_1=F$, $F_2=-D/(4(D-1))F$, $F_3=0$. As a consequence, the linearized equations of the theory impose the condition (\ref{oop}), namely,
\begin{equation}
-\frac{(D-2)}{64\pi G} R^{(1)}=0\, ,
\end{equation}
so the trace of the equation has no dynamics  and one is left with
\begin{equation}
\frac{1}{32\pi G} \left[1+ F(\ell^2 \bar \Box) \ell^2 \bar \Box \right] G_{ab}^{(1)}=0\, .
\end{equation}
By going to the Lorentz gauge as in Section \ref{minko}, one finds
\begin{equation}
-\frac{1}{64\pi G}\left[1+ F\left(\ell^2\bar \Box\right) \ell^2 \bar \Box\right]\bar \Box h_{\langle ab \rangle}\, ,
\end{equation}
and the corresponding propagator is given by
%
\begin{equation}\displaystyle
P_{D}(k)=\frac{64\pi G_{D}}{(D-2) } \left[\frac{i \ell k Y_{\frac{D+2}{2}}(i \ell k)}{Y_{\frac{D}{2}}(i \ell k)}-D\right]^{-1}\, .
\end{equation}
Using this we can analyze the pole structure in various dimensions.

\subsubsection{Three dimensions}
In $D=3$ the propagator becomes
\begin{align}
\frac{P_{3}(k)}{64\pi G_3}=\frac{1}{\ell^2 k^2}-\frac{\ell k \tanh (\ell k)}{\ell^2 k^2}\, .
\end{align}
Studying its pole structure we find a massless mode as well as an infinite tower of massive gravitons. The massless mode is the same as the one appearing in the pure Einstein gravity spectrum and it is pure gauge in three dimensions. On the other hand, the massive gravitons have masses
\begin{equation}
 m_{n}=\frac{\pi}{2\ell }(2n-1)\, ,\quad n= 1,2, \ldots\, ,
\end{equation}
and all of them have negative kinetic energy. This can be seen by expanding the propagator around each of the poles and comparing the overall sign with the one of the positive-energy would-be massless mode. For this, one has
\begin{align}
\frac{P_{3}(k^2\rightarrow 0)}{64\pi G_3}=\frac{1}{\ell^2 k^2}+\mathcal{O}(1) \, .
\end{align}
For the new modes one finds, instead,
\begin{align}
\frac{P_{3}(k^2\rightarrow -m_n^2  )}{64\pi G_3}=-\frac{2}{\ell^2 [k^2+m_n^2]}+\mathcal{O}(1) \, .
\end{align}
Hence, all the new modes are ghosts.

\subsubsection{Four dimensions}
In $D=4$, the analysis of the propagator becomes more cumbersome. To begin with, there is no simplified way to write down the propagator in terms of trigonometric functions. Instead, are left with
\begin{equation}
\frac{P_{4}(k)}{64 \pi G_4}= \frac{i Y_{2}(i \ell k)}{2 \ell k Y_{1}(i \ell k)}\,.
\end{equation}
 Again, we find the Einstein-like massless graviton and an infinite tower of massive ghost gravitons, with masses
 \begin{equation}
 m_n\approx \frac{\pi}{\ell} (0.69937, 1.72832, 2.73619, 3.73987, 4.742, 5.74339, 6.74437, 7.7451,\dots)\,.
 \end{equation}
In this case, the masses are not equispaced, but the difference between pairs of modes tends to $\pi/\ell$ as $n \rightarrow \infty$. Indeed, the $m_n$ tend to $\frac{\pi}{\ell} \left( n - 1/4 \right)$ as $n \rightarrow \infty$. 
Moreover, we now find a tower of modes with complex squared masses which are conjugate of each other,
 \begin{equation}
 m_{n, \rm \pm} \approx \frac{\pi}{\ell} (\pm 0.1790 + 1.220 i, \pm 0.1762 + 2.233 i, \pm 0.1755 + 3.238 i, \cdots) \,.
 \end{equation}
These tend to $\frac{\pi}{\ell} (\pm 0.17485 + (n+1/4) i)$ as $n \rightarrow \infty$.
Again we find that all massive modes, including the complex ones, have negative kinetic energy, namely,
 \begin{align}
\frac{P_{4}(k^2\rightarrow -m_j^2  )}{64\pi G_4}=-\frac{1}{\ell^2 [k^2+m_j^2]}+\mathcal{O}(1) \, ,
\end{align}
$\forall j \in \{n, \pm \}$ so again they are all ghosts. 

\subsubsection{Five dimensions}
In $D=5$ one finds
\begin{equation}
\frac{P_{5}(k)}{64\pi G_5}=\frac{1}{\ell^2 k^2}+\frac{1}{3-3 \ell k \tanh (\ell k)}\, ,
\end{equation}
In addition to the Einstein-like massless graviton, we again find an infinite tower of massive gravitons with masses
 \begin{equation}
 m_n\approx \frac{\pi}{\ell} (0.89075, 1.9485, 2.9660, 3.9746, 4.9797, 5.9831, 6.9855,\dots)\,.
 \end{equation}
Now, however, there is only one tachyonic mode with imaginary mass
 \begin{equation}
 m^2_{\rm t}\approx -\frac{1.43923}{\ell^2}\, .
 \end{equation}
Once again, we find that all the massive modes have negative kinetic energy, namely,
 \begin{align}
\frac{P_{5}(k^2\rightarrow -m_j^2  )}{64\pi G_5}=-\frac{2}{3\ell^2 [k^2+m_j^2]}+\mathcal{O}(1) \, ,
\end{align}
 $\forall j\in \{n, {\rm t}\}$, so they are all ghosts. 

\subsubsection{Six dimensions}
The case of $D=6$ is similar to the four-dimensional case. The propagator reads
\begin{equation}
\frac{P_{6}(k)}{64 \pi G_6}= \frac{i Y_{3}(i \ell k)}{4 \ell k Y_{2}(i \ell k)}\,,
\end{equation}
  and again, we find the Einstein-like massless graviton, an infinite tower of massive ghost gravitons, with masses
 \begin{equation}
 m_n\approx \frac{\pi}{\ell} (1.077, 2.163, 3.191, 4.205, 5.214, 6.220, 7.224,\dots)\,,
 \end{equation}
which tend to $\frac{\pi}{\ell} \left( n + 1/4 \right)$ as $n \rightarrow \infty$; and a tower of modes with complex squared masses which are conjugate of each other,
 \begin{equation}
 m_{n, \rm \pm} \approx \frac{\pi}{\ell} (\pm 0.3382 + 0.4711 i, \pm 0.1877 + 1.636 i, \pm 0.1795 + 2.680 i, \pm 0.1773 + 
 3.699 i, \cdots) \,.
 \end{equation}
These tend to $\frac{\pi}{\ell} (\pm 0.17485 + (n-1/4) i)$ as $n \rightarrow \infty$. Moreover, we find an extra conjugate pair,
 \begin{equation}
 m_{0, \rm \pm} \approx \frac{\pi}{\ell} \pm 0.4716 - 0.1503 i \,.
 \end{equation}
As before, all massive modes are ghosts, including the complex ones, since 
 \begin{align}
\frac{P_{6}(k^2\rightarrow -m_j^2  )}{64\pi G_6}=-\frac{1}{2 \ell^2 [k^2+m_j^2]}+\mathcal{O}(1) \,,
\end{align}
$\forall j \in \{n, \pm \}$, so they all have negative kinetic energy.

\subsubsection{Seven dimensions}
Finally, in $D=7$, one finds
\begin{equation}
\frac{P_{7}(k)}{64 \pi G_7}=\frac{1}{15} + \frac{1}{\ell^2 k^2} - \frac{\ell^2 k^2}{15(3 + \ell^2 k^2 - 3 \ell k \tanh (\ell k))} \, ,
\end{equation}
 Again, we find the Einstein-like massless graviton, and an infinite tower of massive ghost gravitons with masses
 \begin{equation}
 m_n\approx \frac{\pi}{\ell} (1.2604, 2.3719, 3.4109,4.4314, 5.4442,6.4529,7.4593,\dots)
 \end{equation}
 with the difference between pairs of modes tending to $\pi/\ell$ as $n \rightarrow \infty$. Now, there are only two extra modes with complex squared masses which are conjugate of each other, namely,
 \begin{equation}
 m^2_{\rm \pm}\approx -\frac{2.01933 \pm 3.19512 i}{\ell^2}\, .
 \end{equation} 
Once more, we find that all the massive modes, including the ones with complex squared-masses, have negative kinetic energy, namely,
 \begin{align}
\frac{P_{7}(k^2\rightarrow -m_j^2  )}{64\pi G_7}=-\frac{2}{5\ell^2 [k^2+m_j^2]}+\mathcal{O}(1) \, ,
\end{align}
$\forall j \in \{n, \pm \}$ so they are all ghosts.  

We have found that, regardless of the number of dimensions, there are always pathological modes appearing in the linearized spectrum of these brane-world gravities, with squared masses of order $\sim 1/\ell^2$.
Since the bulk theory is Einstein gravity, which is perfectly well-defined, the appearance of these pathological modes on the gravitational effective theory induced on the brane might seem worrisome at first. The bulk, however, is dual to this induced theory on the brane plus a cut-off CFT, which we have neglected in this analysis. The CFT cut-off is precisely $\sim 1/\ell^2$, and so it is not surprising that pathologies might appear at this order.
Moreover, when one takes the coupling between this cut-off CFT and the induced gravity on the brane into account, the observed pathologies disappear. In a sense, coupling the induced action to the cut-off CFT allows one to ``UV-complete'' the theory by making it dual to the perfectly defined Einstein gravity in the bulk. These results, along with a careful analysis of the linear spectrum in this case, will appear in future work.



\section{Conclusions}\label{conclu}
A summary of the main findings of this paper can be found in the introduction. Let us close with some comments regarding open questions and future work. 

In this work we have initiated the study of GQTs with covariant derivatives. Our analysis has been restricted to four dimensions and to the first few curvature orders. It would be interesting to pursue a full classification of GQTs with covariant derivatives in general dimensions as well as for arbitrary curvature orders, similar to the one achieved for polynomial GQTs in \cite{Bueno:2022res,Moreno:2023rfl}. Similarly, it would be interesting to determine whether the departure from the universal behavior observed in polynomial GQTs for the temperature of small black holes in the case of the new GQT with covariant derivatives extends to other theories of that kind, and whether a new universal behavior arises in that case. The implications for the evaporation process of black holes should also be studied in this context. 

Additionally, it would be interesting to prove that any gravitational effective action can be mapped to a GQT. This is established for general polynomial densities \cite{Bueno:2019ltp}, but the proof for terms involving covariant derivatives is thus far limited to theories with up to eight derivatives  of  the metric and also for theories with any number of Riemann tensors and two covariant derivatives.

On a different front, it would be interesting to characterise the generalized symmetries of general linearized higher-curvature gravities with covariant derivatives along the lines of \cite{Benedetti:2023ipt}, where such analysis was performed for $\mathcal{L}(g^{ab},R_{abcd})$ theories.

Regarding brane-world gravities, the existence of ghosts as well as of tachyonic and complex-squared-mass modes in the linearized spectrum of these effective theories seems to be in tension with the absence of such pathologies in the bulk theory (Einstein gravity).  In particular, the appearance of imaginary poles in the propagators of particles has been suggested as an indication of confinement ---see \eg \cite{Stainsby:1990fh,Maris:1995ns,Hayashi:2021nnj}. 
It would be interesting to understand their origin from the bulk perspective. Naturally, here we have ignored the effects of the cut-off CFT which is also induced on the brane, so one could try to understand if and how its coupling to the brane-world gravities resolves the pathologies.

\section*{Acknowledgements}
We thank Roberto Emparan for useful discussions. We thank Ivan Kol{\'a}{\v r} for a number of helpful comments on the first version of the manuscript. 
SEAG thanks the University of Amsterdam for hospitality during the completion of this work, and acknowledges mobility support from the Research Foundation - Flanders (FWO). The work of SEAG is partially supported by the FWO Research Project G0H9318N and the inter-university project iBOF/21/084. The work of PB was supported by a Ram\'on y Cajal fellowship (RYC2020-028756-I) from Spain’s Ministry of Science and Innovation. PAC was partially supported by a postdoctoral fellowship from the Research Foundation-Flanders (FWO grant 12ZH121N). The project that gave rise to these results received the support of a fellowship from ``la Caixa'' Foundation (ID 100010434) with code LCF/BQ/PI23/11970032. RAH is grateful to the Department of Applied Mathematics and Theoretical Physics at the University of Cambridge for hospitality during the completion of this work. The work of RAH received the support of a fellowship from ``laCaixa'' Foundation (ID 100010434) and from the European Union’s Horizon 2020 research and innovation programme under the Marie Sklodowska-Curie grant agreement No 847648 under fellowship code LCF/BQ/PI21/11830027. The work of QL is supported by the Spanish Ministry of Universities through FPU grant No. FPU19/04859.  PB, PAC, RAH and QL also acknowledge financial support from MICINN grant PID2019-105614GB-C22, AGAUR grant 2017-SGR 754, and State Research Agency of MICINN through the `Unit of Excellence María de Maeztu 2020-2023' award to the Institute of Cosmos Sciences (CEX2019-000918-M).

\appendix

\section{Basis of invariants}\label{basis}

We present here a complete list of the curvature invariants at each order in derivatives. The same list can be found in~\cite{Fulling:1992vm}. Our ordering also follows~\cite{Fulling:1992vm}: The invariants are ordered by the number of covariant derivatives acting on individual curvature tensors. We begin with those invariants that involve the largest number of derivatives acting on curvature, and end with the polynomial curvature invariants (those built exclusively from contractions of the Riemann tensor). 

\subsection{Four derivatives}
There are four possible terms involving four derivatives of the metric:
\begin{align}
\mathcal{R}_4^{(1)} &= \Box R \, , 
\quad 
\mathcal{R}_4^{(2)} = R^2 \, ,
\quad 
\mathcal{R}_4^{(3)} = R^{pq} R_{pq} \, ,
\quad
\mathcal{R}_4^{(4)} = R^{pqrs} R_{pqrs} \, .
\end{align}

\subsection{Six derivatives}
There are 17 terms involving six derivatives of the metric:
\begin{align}
\mathcal{R}_{6}^{(1)} &= \Box R^2 \, ,
\quad
\mathcal{R}_{6}^{(2)} = R \Box R \, ,
\quad
\mathcal{R}_{6}^{(3)} = R^{;pq} R_{;pq} \, ,
\quad 
\mathcal{R}_{6}^{(4)} = R^{pq} \Box R_{pq} \, ,
\quad 
\mathcal{R}_{6}^{(5)} = R^{pq;rs} R_{pqrs} \, ,
\nonumber\\
\mathcal{R}_{6}^{(6)} &= R^{;p} R_{;p} \, ,
\quad 
\mathcal{R}_{6}^{(7)} = R^{pq;r} R_{pq;r} \, ,
\quad
\mathcal{R}_{6}^{(8)} = R^{pq;r} R_{pr;q} \, ,
\quad 
\mathcal{R}_{6}^{(9)} = R^{pqrs;t} R_{pqrs;t} \, ,
\quad 
\mathcal{R}_{6}^{(10)} = R^3 \, ,
\nonumber\\
\mathcal{R}_{6}^{(11)} &= R R^{pq} R_{pq} \,,
\quad
\mathcal{R}_{6}^{(12)} = R^{pq} \tensor{R}{_p^r} R_{qr} \, ,
\quad 
\mathcal{R}_{6}^{(13)} = R^{pq} R^{rs} R_{prqs} \, ,
\quad
\mathcal{R}_{6}^{(14)} = R R^{pqrs} R_{pqrs} \, ,
\nonumber\\
\mathcal{R}_{6}^{(15)} &= R^{pq} \tensor{R}{^{rst}_p} R_{rstq} \, ,
\quad
\mathcal{R}_{6}^{(16)} = R^{pqrs} \tensor{R}{_{pq}^{tu}} R_{rstu} \, ,
\quad
\mathcal{R}_{6}^{(17)} = R^{pqrs} \tensor{R}{_p^t_r^u} R_{qtsu} \, .
\end{align}

\subsection{Eight derivatives}
There are 92 terms involving eight derivatives of the metric:
\begin{align}
\mathcal{R}_{8}^{(1)} &= \Box^3 R \, ,
\quad
\mathcal{R}_{8}^{(2)} = R \Box^2 R \, ,
\quad
\mathcal{R}_{8}^{(3)} = R_{pq} \Box\tensor{R}{^{;pq}} \,,
\quad
\mathcal{R}_{8}^{(4)} = R^{pq} \Box^2 \tensor{R}{_{pq}} \, ,
\quad
\mathcal{R}_{8}^{(5)} = \tensor{R}{^{pq;rs}} \tensor{R}{_{prqs}}
\, ,
\nonumber\\
\mathcal{R}_{8}^{(6)} &= \tensor{R}{^{;p}} \Box \tensor{R}{_{;p}} \, ,
\quad
\mathcal{R}_{8}^{(7)} = \tensor{R}{^{pq;r}} \tensor{R}{_{pq;r}} \, ,
\quad
\mathcal{R}_{8}^{(8)} = \tensor{R}{^{pq;r}} \Box \tensor{R}{_{pq;r}} \, ,
\quad
\mathcal{R}_{8}^{(9)} = \tensor{R}{^{pq;r}} \Box \tensor{R}{_{pr;q}} \, ,
\nonumber\\
\mathcal{R}_{8}^{(10)} &= \tensor{R}{^{pq;rst}} \tensor{R}{_{prqs;t}} \, ,
\quad
\mathcal{R}_{8}^{(11)} = \left( \Box R \right)^2 \, ,
\quad
\mathcal{R}_{8}^{(12)} = \tensor{R}{^{;pq}} \tensor{R}{_{;pq}} \, ,
\quad
\mathcal{R}_{8}^{(13)} = \tensor{R}{^{;pq}} \Box \tensor{R}{_{pq}} \, ,
\nonumber\\
\mathcal{R}_{8}^{(14)} &= \Box R^{pq} \Box R_{pq} \, , 
\quad
\mathcal{R}_{8}^{(15)} = \tensor{R}{^{pq;rs}} \tensor{R}{_{pq;rs}} \, ,
\quad
\mathcal{R}_{8}^{(16)} = \tensor{R}{^{pq;rs}} \tensor{R}{_{pr;qs}} \, ,
\quad
\mathcal{R}_{8}^{(17)} = \tensor{R}{^{pq;rs}} \tensor{R}{_{rs;pq}} \, ,
\nonumber\\
\mathcal{R}_{8}^{(18)} &= \tensor{R}{^{pqrs;tu}} \tensor{R}{_{pqrs;tu}} \, ,
\quad
\mathcal{R}_{8}^{(19)} = R^2 \Box R \, , 
\quad 
\mathcal{R}_{8}^{(20)} = R \tensor{R}{^{;pq}} R_{pq} \, ,
\quad 
\mathcal{R}_{8}^{(21)} = \Box R R^{pq} R_{pq} \, ,
\nonumber\\
\mathcal{R}_{8}^{(22)} &= R R^{pq} \Box R_{pq} \, ,
\quad
\mathcal{R}_{8}^{(23)} = R^{;pq} \tensor{R}{_p^r} R_{qr} \, ,
\quad
\mathcal{R}_{8}^{(24)} = R^{pq}  \tensor{R}{_p^r} \Box R_{qr} \, ,
\quad 
\mathcal{R}_{8}^{(25)} = R^{pq} R^{rs} R_{pq;rs} \, ,
\nonumber\\
\mathcal{R}_{8}^{(26)} &= R^{pq} R^{rs} R_{pr;qs} \, ,
\quad
\mathcal{R}_{8}^{(27)} = R^{;pq} R^{rs} R_{prqs} \, ,
\quad 
\mathcal{R}_{8}^{(28)} = R R^{pq;rs} R_{prqs} \, ,
\quad 
\mathcal{R}_{8}^{(29)} = R^{pq} \Box R^{rs} R_{prqs} \, ,
\nonumber\\
\mathcal{R}_{8}^{(30)} &= R^{pq} \tensor{R}{_p^{r;st}} R_{qsrt} \, ,
\quad 
\mathcal{R}_{8}^{(31)} = R^{pq} \tensor{R}{^{rs}_{;q}^t} R_{prst} \, ,
\quad 
\mathcal{R}_{8}^{(32)} = \Box R R^{pqrs} R_{pqrs} \, ,
\quad 
\mathcal{R}_{8}^{(33)} = R^{;pq} \tensor{R}{^{rst}_q} R_{rstp} \, ,
\nonumber\\
\mathcal{R}_{8}^{(34)} &= \Box R^{pq} \tensor{R}{^{rst}_p} R_{rstq} \, ,
\quad 
\mathcal{R}_{8}^{(35)} = R^{pq;rs} \tensor{R}{^{tu}_{pr}} R_{tuqs} \, ,
\quad
\mathcal{R}_{8}^{(36)} = R^{;pqrs} \tensor{R}{^t_p^u_q}R_{trus} \, , 
\nonumber\\
\mathcal{R}_{8}^{(37)} &= R^{pq;rs} \tensor{R}{^t_p^u_r} R_{tqus} \, ,
\quad
\mathcal{R}_{8}^{(38)} = R^{pq} R^{rstu} R_{rstu;pq} \, ,
\quad
\mathcal{R}_{8}^{(39)} = R^{pqrs} \tensor{R}{_p^{tuv}} R_{qtru;sv} \, ,
\nonumber\\
\mathcal{R}_{8}^{(40)} &= R R^{;p} R_{;p} \, , 
\quad 
\mathcal{R}_{8}^{(41)} = R^{;p} R^{;q} R_{pq} \, ,
\quad
\mathcal{R}_{8}^{(42)} = R R^{pq;r} R_{pq;r} \, ,
\quad
\mathcal{R}_{8}^{(43)} = R R^{pq;r} R_{pr;q} \, ,
\nonumber\\
\mathcal{R}_{8}^{(44)} &= R^{;p} R^{qr} R_{qr;p} \, ,
\quad
\mathcal{R}_{8}^{(45)} = R^{;p} R^{qr} R_{pq;r} \, ,
\quad 
\mathcal{R}_{8}^{(46)} = R^{pq} \tensor{R}{_p^{r;s}} R_{qr;s} \,,
\quad 
\mathcal{R}_{8}^{(47)} = R^{pq} \tensor{R}{_p^{r;s}} R_{qs;r} \, ,
\nonumber\\
\mathcal{R}_{8}^{(48)} &=  R{^pq} \tensor{R}{^{rs}_{;p}} R_{rs;q} \, , 
\quad 
\mathcal{R}_{8}^{(49)} = R^{pq} \tensor{R}{^{rs}_{;p}} R_{rq;s} \, , 
\quad
\mathcal{R}_{8}^{(50)} = R^{;p} R^{qr;s} R_{pqrs} \, ,
\nonumber\\
\mathcal{R}_{8}^{(51)} &=   R^{pq;r} \tensor{R}{_p^{s;t}} R_{qrst} \, ,
\quad
\mathcal{R}_{8}^{(52)} = R^{pq;r} \tensor{R}{_p^{s;t}} R_{qsrt} \, ,
\quad
\mathcal{R}_{8}^{(53)} = R^{qr;p} \tensor{R}{^{st}_{;p}} R_{qsrt} \,,
\nonumber\\
\mathcal{R}_{8}^{(54)} &= R^{pq;r} \tensor{R}{^{st}_{;p}} \tensor{R}{_{qsrt}}  \, , 
\quad
\mathcal{R}_{8}^{(55)} =  R^{pq} R^{rs;t} R_{prqs;t} \, ,
\quad
\mathcal{R}_{8}^{(56)} =  R^{pq}R^{rs;t} R_{rtsp;q} \, ,
\nonumber\\
\mathcal{R}_{8}^{(57)} &=  R^{;p} R^{qrst} R_{qrst;p} \, ,
\quad
\mathcal{R}_{8}^{(58)} = R R^{pqrs;t} R_{pqrs;t} \, ,
\quad 
\mathcal{R}_{8}^{(59)} = R^{pq} \tensor{R}{^{rstu}_{;p}} R_{rstu;q} \, ,
\nonumber\\
\mathcal{R}_{8}^{(60)} &= R^{pq} \tensor{R}{^{rstu}_{;p}} R_{rstq;u} \, ,
\quad
\mathcal{R}_{8}^{(61)} = R^{pq;r} \tensor{R}{^{st}_r^u} R_{stpq;u} \, ,
\quad
\mathcal{R}_{8}^{(62)} = R^{pq;r} \tensor{R}{^{st}_p^u} R_{stqr;u} \, ,
\nonumber\\
\mathcal{R}_{8}^{(63)} &= R^{pq;r} \tensor{R}{^s_p^{tu}} R_{sqtr;u} \, ,
\quad
\mathcal{R}_{8}^{(64)} = R^{pqrs} \tensor{R}{^{tuv}_{p;q}} R_{tuvr;s} \, ,
\quad
\mathcal{R}_{8}^{(65)} = R^{pqrs} \tensor{R}{^{tuv}_{p;r}} R_{tuvq;s} \, ,
\nonumber\\
\mathcal{R}_{8}^{(66)} &= R^{pqrs} \tensor{R}{^t_p^u_r^{;v}} R_{tqus;v} \, ,
\quad 
\mathcal{R}_{8}^{(67)} = R^4 \, ,
\quad 
\mathcal{R}_{8}^{(68)} = R^2 R^{pq} R_{pq} \, ,
\quad
\mathcal{R}_{8}^{(69)} = R R^{pq} \tensor{R}{_p^r} R_{qr} \, ,
\nonumber\\
\mathcal{R}_{8}^{(70)} &= \left( R^{pq} R_{pq} \right)^2 \, ,
\quad
\mathcal{R}_{8}^{(71)} = R^{pq} \tensor{R}{_p^r} \tensor{R}{_q^s} R_{rs} \, ,
\quad
\mathcal{R}_{8}^{(72)} = R R^{pq} R^{rs} R_{prqs} \, ,
\quad
\mathcal{R}_{8}^{(73)} = R^{pq} R^{rs} \tensor{R}{_r^t} R_{psqt} \, ,
\nonumber\\
\mathcal{R}_{8}^{(74)} &= R^2 R^{pqrs} R_{pqrs} \, ,
\quad
\mathcal{R}_{8}^{(75)} = R R^{pq} \tensor{R}{^{rst}_p} R_{rstq} \, ,
\quad
\mathcal{R}_{8}^{(76)} = R^{pq} R_{pq} R^{rstu} R_{rstu} \, ,
\nonumber\\
\mathcal{R}_{8}^{(77)} &= R^{pq} \tensor{R}{_p^r} \tensor{R}{^{stu}_q} R_{stur} \, ,
\quad 
\mathcal{R}_{8}^{(78)} = R^{pq} R^{rs} \tensor{R}{^{tu}_{pr}} R_{tuqs} \, , 
\quad
\mathcal{R}_{8}^{(79)} = R^{pq} R^{rs} \tensor{R}{^t_p^u_q} R_{trus} \, ,
\nonumber\\
\mathcal{R}_{8}^{(80)} &= R^{pq} R^{rs} \tensor{R}{^t_p^u_r} R_{tqus} \, ,
\quad
\mathcal{R}_{8}^{(81)} = R R^{psrs} \tensor{R}{_{pq}^{tu}} R_{rstu} \, ,
\quad
\mathcal{R}_{8}^{(82)} = R R^{pqrs} \tensor{R}{_p^t_r^u} R_{qtsu} \, ,
\nonumber\\
\mathcal{R}_{8}^{(83)} &= R^{pq} \tensor{R}{_p^r_q^s} \tensor{R}{^{tuv}_r} \tensor{R}{_{tuvs}} \, ,
\quad
\mathcal{R}_{8}^{(84)} = R^{pq} R^{rstu} \tensor{R}{_{rs}^v_p}R_{tuvq} \, ,
\quad 
\mathcal{R}_{8}^{(85)} = R^{pq} R^{rstu} \tensor{R}{_r^v_{tp}} R_{svuq} \, ,
\nonumber\\
\mathcal{R}_{8}^{(86)} &= \left( R^{pqrs} R_{pqrs} \right)^2 \, ,
\quad 
\mathcal{R}_{8}^{(87)} = R^{pqrs} \tensor{R}{_{pq}^{tu}} \tensor{R}{_{tu}^{vw}} R_{rsvw} \, ,
\quad
\mathcal{R}_{8}^{(88)} = R^{pqrs} \tensor{R}{_{pq}^{tu}} \tensor{R}{_{tu}^{vw}} R_{rsvw} \, ,
\nonumber\\
\mathcal{R}_{8}^{(89)} &= R^{pqrs} \tensor{R}{_{pq}^{tu}} \tensor{R}{_{rt}^{vw}} R_{suvw} \, ,
\quad
\mathcal{R}_{8}^{(90)} = R^{pqrs} \tensor{R}{_{pq}^{tu}} \tensor{R}{_r^v_t^w} R_{svuw} \, ,
\quad 
\mathcal{R}_{8}^{(91)} = R^{pqrs} \tensor{R}{_p^t_r^u} \tensor{R}{_t^v_u^w} R_{qvsw} \, ,
\nonumber\\
\mathcal{R}_{8}^{(92)} &= R^{pqrs} \tensor{R}{_p^t_r^u} \tensor{R}{_t^v_q^w} R_{uvsw}  \, .
\end{align}

\subsection{Ten derivatives}
The number of independent invariants grows rapidly with an increasing number of derivatives.
To the best of our knowledge, a complete classification of terms involving more than eight-derivatives of the metric has not been completed. However, for example, at ten-derivative order it is known that there are 668 invariants. The set of ten-derivative invariants we have used consists of $180 = 20 + 92 + 4\times 17$ elements, and so it is necessarily \textit{very} incomplete. Out of the $180$ densities that we use, only 20 are not built from products of lower-order densities. These are 
\begin{align}
\nonumber
\mathcal{R}_{10}^{(1)} &= \tensor{C}{_{abcd}} \tensor{C}{^{abcd}}  \tensor{C}{^{efgh;i}} \tensor{C}{_{efgh;i}}\, 
\nonumber\\
\mathcal{R}_{10}^{(2)} &= R_b^a R_d^b R_f^c \tensor{R}{_{ag}^{de}} \tensor{R}{_{ce}^{fg}} \, ,
\quad
\mathcal{R}_{10}^{(3)}  = R_b^a R_d^b R_f^c \tensor{R}{_{cg}^{de}} \tensor{R}{_{ae}^{fg}} \, ,
\quad 
\mathcal{R}_{10}^{(4)}  = R_b^a R_c ^b \tensor{R}{_{ae}^{cd}} \tensor{R}{_{gh}^{ef}} \tensor{R}{_{df}^{gh}} \, ,
\nonumber
\\
\mathcal{R}_{10}^{(5)} &= R_b^a R_c^b \tensor{R}{_{ef}^{cd}} \tensor{R}{_{gh}^{ef}} \tensor{R}{_{ad}^{gh}} \, ,
\quad
\mathcal{R}_{10}^{(6)} = R_b^a R_c^b \tensor{R}{_{eg}^{cd}} \tensor{R}{_{ah}^{ef}} \tensor{R}{_{df}^{gh}} \, ,
\quad
\mathcal{R}_{10}^{(7)} = R_c^a R_d^b \tensor{R}{_{ab}^{cd}}\tensor{R}{_{gh}^{ef}}\tensor{R}{_{ef}^{gh}} \, ,
\nonumber\\
\mathcal{R}_{10}^{(8)} &= R_c^a R_d^b \tensor{R}{_{ae}^{cd}}\tensor{R}{_{gh}^{ef}}\tensor{R}{_{bf}^{gh}} \, ,
\quad 
\mathcal{R}_{10}^{(9)} = R_c^a R_d ^b \tensor{R}{_{ef}^{cd}}\tensor{R}{_{gh}^{ef}} \tensor{R}{_{ab}^{gh}} \, ,
\quad 
\mathcal{R}_{10}^{(10)} = R_c^a R_d^b \tensor{R}{_{eg}^{cd}} \tensor{R}{_{ah}^{ef}} \tensor{R}{_{bf}^{gh}} \, ,
\nonumber\\
\mathcal{R}_{10}^{(11)} &= R_c^a R_e^b \tensor{R}{_{af}^{cd}} \tensor{R}{_{gh}^{ef}} \tensor{R}{_{bd}^{gh}} \, ,
\quad
\mathcal{R}_{10}^{(12)} = R_b^a \tensor{R}{_{ad}^{bc}} \tensor{R}{_{fh}^{de}}\tensor{R}{_{c i}^{fg}} \tensor{R}{_{eg}^{hi}} \, ,
\quad
\mathcal{R}_{10}^{(13)} = R_b^a \tensor{R}{_{de}^{bc}}\tensor{R}{_{cf}^{de}} \tensor{R}{_{hi}^{fg}} \tensor{R}{_{ag}^{hi}} \, ,
\nonumber\\
\mathcal{R}_{10}^{(14)} &= R_b^a \tensor{R}{_{df}^{bc}} \tensor{R}{_{ac}^{de}} \tensor{R}{_{hi}^{fg}} \tensor{R}{_{eg}^{hi}} \, ,
\quad 
\mathcal{R}_{10}^{(15)} = R_b^a \tensor{R}{_{df}^{bc}}\tensor{R}{_{ah}^{de}}\tensor{R}{_{ei}^{fg}}\tensor{R}{_{cg}^{hi}} \, ,
\quad 
\mathcal{R}_{10}^{(16)} = R_b^a \tensor{R}{_{df}^{bc}} \tensor{R}{_{gh}^{de}} \tensor{R}{_{ei}^{fg}} \tensor{R}{_{ac}^{hi}} \, ,
\nonumber\\
\mathcal{R}_{10}^{(17)} &= \tensor{R}{_{cd}^{ab}} \tensor{R}{_{eg}^{cd}} \tensor{R}{_{ai}^{ef}} \tensor{R}{_{fj}^{gh}} \tensor{R}{_{bh}^{ij}} \, ,
\quad
\mathcal{R}_{10}^{(18)} = \tensor{R}{_{ce}^{ab}} \tensor{R}{_{af}^{cd}} \tensor{R}{_{gi}^{ef}} \tensor{R}{_{bj}^{gh}} \tensor{R}{_{dh}^{ij}} \, ,
\nonumber\\
\mathcal{R}_{10}^{(19)} &= \tensor{R}{_{ce}^{ab}} \tensor{R}{_{ag}^{cd}} \tensor{R}{_{bi}^{ef}} \tensor{R}{_{fj}^{gh}} \tensor{R}{_{dh}^{ij}} \, ,
\quad
\mathcal{R}_{10}^{(20)} = \tensor{R}{_{ce}^{ab}} \tensor{R}{_{fg}^{cd}} \tensor{R}{_{hi}^{ef}} \tensor{R}{_{aj}^{gh}} \tensor{R}{_{bd}^{ij}} \,  .
\end{align}


\section{Hairy black holes in pure eight-derivative GQTs}\label{8dpure}

Excluding the fourth-order density $\mathcal{L}_{(8,0)}^{(1)}$, all the other eight-derivative Lagrangians allow the Schwarzschild-(A)dS spacetime as an exact solution. However, due to the fact that the equations of motion are of higher order, one may wonder if additional solutions exist.

In order to illustrate the possibility of having non-Schwarzschild black holes, let us  consider the simple (yet unrealistic) case in which we do not have an Einstein-Hilbert term in the action. In fact, for the sake of simplicity let us just consider a higher-derivative gravity given by the following eight-derivative Lagrangian,
\begin{equation}
\mathcal{L}=\frac{1}{16\pi G} \left[ c_{(2)}\mathcal{L}_{(8,2)}^{(2)}+c_{(3)}\mathcal{L}_{(8,4)}^{(3)}+c_{(4)}\mathcal{L}_{(8,4)}^{(4)}\right]\, .
\end{equation}
The integrated equation for the function $f$ becomes in this case 
\begin{equation}
c_{(2)}\mathcal{F}_{(8,2)}^{(2)}+c_{(3)}\mathcal{F}_{(8,4)}^{(3)}+c_{(4)}\mathcal{F}_{(8,4)}^{(4)}=4GM\, ,
\end{equation}
which explicitly reads
\begin{equation}
\frac{f^2}{2 r^5}\left(\alpha ^2 \left(-8 c_{(2)}+15 c_{(3)}-16 c_{(4)}\right)+3 c_{(3)} r^2 \alpha'^2+2 \alpha  r \left(2 c_{(4)} r \alpha ''-3 c_{(3)} \alpha '\right)\right)=4GM\, .
\end{equation}
Now, for $M=0$, a solution to this equation is the Schwarzschild-(A)dS black hole with arbitrary mass parameter and cosmological constant, which has $\alpha=0$. However, this is not the most general solution. If we again set $M=0$, we find a homogeneous equation for $\alpha$, that has the following general solution
\begin{equation}
\alpha(r)=a r^{\nu } \left(r^{\mu }+b \right){}^{\frac{2-2 \nu }{\mu +1}}\, ,
\end{equation}
where $a$ and $b$ are integration constants, and
\begin{equation}
\label{eq:mu}
\mu\equiv \sqrt{17+8 \frac{c_{(2)}}{c_{(4)}}+\frac{6c_{(2)}c_{(3)}-9c_{(3)}^2}{c_{(4)}^2}}\, ,\quad
\nu\equiv \frac{3 \frac{c_{(3)}}{c_{(4)}}-2 \mu +2}{3 \frac{c_{(3)}}{c_{(4)}}+4}\, .
\end{equation}
Thus, solving now the equation
\begin{equation}
r^2 f''+2 (k-f)=\alpha\, ,
\end{equation}
we get the general solution
\begin{equation}
\begin{aligned}
f(r)=&-\lambda r^2+k-\frac{2m}{r}\\
&+\frac{a r^{\nu } \left(b+r^{\mu }\right)^{\frac{3+\mu -2 \nu
   }{1+\mu }}}{3 b} \left[\frac{1}{\nu-2} \, _2F_1\left(1,\frac{(\mu-1) (2+\mu -\nu )}{\mu  (1+\mu
   )};\frac{\mu +\nu-2 }{\mu };-\frac{r^{\mu }}{b}\right)\right.\\
   &\left.-\frac{1}{\nu+1} \,
   _2F_1\left(1,\frac{1+4 \mu +\mu ^2+\nu -\mu  \nu }{\mu( 1+\mu )};\frac{1+\mu +\nu
   }{\mu };-\frac{r^{\mu }}{b}\right)\right]\, ,
\end{aligned}
\end{equation}
where $\lambda$ and $m$ arise as integration constants. This represents a biparametric modification of Schwarzschild's solution. In the limit $b\rightarrow 0$, we get a much simpler solution,
\begin{equation}
f(r)=-\lambda r^2+k-\frac{2m}{r}+\tilde a \, r^{\sigma}\, ,\qquad \sigma=\frac{3 c_{(3)}+2c_{(4)} \left(\mu+1\right)}{3 c_{(3)}+4 c_{(4)}}\, ,
\end{equation}
where $\tilde a\propto a$.  For $\sigma<-1$ this represents an asymptotically flat/AdS/dS black hole solution with continuous hair. On the other hand, for $\sigma>2$ the asymptotic behavior is exotic. 

\section{Linearized equations on an AdS background}\label{adsbck}
Here we present the explicit linearized euqations of motion around an AdS background for the simplest examples of the theories considered in the main text.  First consider the effective quadratic theory arising from purely polynomial theories,
\be 
\mathcal{L}_{(0)} = \lambda \left(R - 2 \Lambda_0\right) + \alpha_{(0)} R^2 +\beta_{(0)}  R_{ab} R^{ab}+\gamma_{(0)} R_{abcd} R^{abcd} \, .
\ee
For this theory, the linearized equations were computed in~\cite{Aspects} and read
\begin{align}
 \mathcal{E}_{ab}^{(1)}=&  \left[\frac{\lambda}{2} - \frac{2}{\ell_\star^2}\left( D(D-1) \alpha_{(0)} + (D-2) \beta_{(0)} + (D-3)(D-4)\gamma_{(0)} \right)  + \beta_{(0)} \bar{\Box} \right] G_{ab}^{(1)} 
 \nonumber
\\
&+ \left[2\alpha_{(0)} + \beta_{(0)} \right] \left[\bar{g}_{ab} \bar{\Box} - \bar{\nabla}_a \bar{\nabla}_{b} \right] R^{(1)}  
- \frac{1}{\ell_\star^2} \left[2(D-1)\alpha_{(0)} + \beta_{(0)} \right] \bar{g}_{ab} R^{(1)} \, ,
\end{align}
and we are using the following conventions for the background curvature tensor,
\be 
\bar{R}_{abcd} = - \frac{1}{\ell_\star^2} \left[\bar{g}_{a c} \bar{g}_{bd} - \bar{g}_{a d} \bar{g}_{b c} \right] \, .
\ee
Now consider the effective action involving one d'Alembertian acting on curvature,
\be 
\mathcal{L}_{(1)} =\alpha_{(1)} R\Box  R+\beta_{(1)}  R_{ab}\Box R^{ab}+\gamma_{(1)} R_{abcd} \Box R^{abcd} \, .
\ee
For this theory, the linearized equations of motion take the form
\begin{align}
\mathcal{E}_{ab}^{(1)}=& \left[ -\frac{8}{\ell_\star^4}\gamma_{(1)}  (D-3) D  + \frac{2}{\ell_\star^2} (\beta_{(1)} + 2 \gamma_{(1)} (5 - D) ) \bar{\Box} + (\beta_{(1)} +4 \gamma_{(1)} ) \bar{\Box}^2 \right] G_{ab}^{(1)}
\nonumber
\\
&   
+ \left(2 \alpha_{(1)} + \beta_{(1)} + 2 \gamma_{(1)} \right) \left[\bar{g}_{ab} \bar{\Box} - \bar{\nabla}_a \bar{\nabla}_b \right] \bar{\Box} R^{(1)} + \frac{2 (D-2) (4 \gamma_{(1)} + \beta_{(1)})}{\ell_\star^2} \bar{\nabla}_a \bar{\nabla}_b R^{(1)}
\nonumber 
\\
&- \frac{(D-1)(2 \alpha_{(1)} + \beta_{(1)} + 2 \gamma_{(1)})}{\ell_\star^2}  \bar{g}_{ab} \bar{\Box} R^{(1)} - \frac{4(D-2)(D-3)}{\ell_\star^2} \gamma_{(1)}  \bar{g}_{ab} R^{(1)} \, .
\end{align}
In particular, note that the linearized field equations of this six-derivative action involves terms with two, four, and six derivatives. It may be that the linearized equations can be simplified in alternative gauges, e.g.,~\cite{Kolar:2023mkw}. 

\bibliographystyle{JHEP}
\bibliography{Gravities}

\end{document}